\pdfoutput=1
\documentclass[aps,prb,nolongbibliography,notitlepage,twocolumn,superscriptaddress]{revtex4-2}

\usepackage{amsmath}
\usepackage{amssymb}
\usepackage{bbold}
\usepackage{color}
\usepackage{graphicx}
\graphicspath{{figs/}}
\usepackage[colorlinks=true,linkcolor=black,citecolor=blue,pdfencoding=auto]{hyperref}

\newcommand{\xiloc}{\xi_{\rm loc}}

\begin{document}

\title{Transverse momentum transport and collective-mode localization in dipolar tube arrays}

\author{Camilla Polvara}
\affiliation{D\'epartement de physique, Universit\'e de Montr\'eal, Montr\'eal, Qu\'ebec, Canada H3C 3J7}
\author{Nishchhal Verma}
\affiliation{Department of Physics, Columbia University, New York, New York 10027, USA}
\author{Sarang Gopalakrishnan}
\affiliation{Department of Electrical and Computer Engineering, Princeton University, Princeton, New Jersey 08544, USA}
\author{Vadim Oganesyan}
\affiliation{Physics Program and Initiative for the Theoretical Sciences, The Graduate Center,
CUNY, New York, New York 10016, USA;
Department of Physics and Astronomy, College of Staten Island, CUNY,
Staten Island, New York 10314, USA}

\begin{abstract}
We study transverse transport of longitudinal momentum in an array of one-dimensional Bose gases coupled by dipolar interactions without particle tunneling. Within zero-temperature Bogoliubov theory, we characterize this transport using the inter-tube momentum correlator and its squared magnitude, termed amplitude and intensity. In a uniform array, pairs of Bogoliubov quasiparticles propagate ballistically across the tubes, while their dephasing produces a \(t^{-3}\) decay of the long-wavelength amplitude. Tube-to-tube density imbalance acts as quenched disorder and localizes the transverse collective modes. At fixed longitudinal momentum, this arrests the growth of the intensity profile, whose time-averaged spatial decay measures the localization length. Equal-time correlations are much less sensitive to disorder and primarily reflect the interaction range: in the uniform array they decay with tube separation as \(r^{-7}\) for dipolar coupling, compared with exponential decay for nearest-neighbor coupling. Finally, sufficiently strong inter-tube repulsion destabilizes the longitudinally uniform state toward density modulations staggered between neighboring tubes, at a threshold independent of density imbalance.
\end{abstract}

\maketitle

\section{Introduction}
\label{sec:intro}

Two-dimensional optical lattices split a quantum gas into an array of parallel one-dimensional tubes, the standard setting for one-dimensional Bose gases~\cite{stoferle2004,cazalilla2011}, from the quantum Newton's cradle~\cite{kinoshita2006} to quantitative tests of generalized hydrodynamics~\cite{malvania2021} and the observation of hydrodynamization~\cite{le2023}. With magnetic lanthanides~\cite{lu2011,aikawa2012} the same geometry acquires a long-range dipolar interaction~\cite{lahaye2009,chomaz2023}, which breaks integrability by a tunable amount~\cite{tang2018,li2023} and stabilizes strongly correlated states against collapse~\cite{kao2021}. The dipolar interaction also acts between tubes. Single-particle tunneling falls exponentially with lattice depth; the dipolar coupling falls only as the inverse cube of the separation. There is therefore a regime in which the tubes exchange energy and momentum but not particles, as observed in dysprosium bilayers through sympathetic cooling and coupled collective oscillations~\cite{du2024}. Each of the $N_t$ tubes then keeps its own particle number and its own condensate phase, a $U(1)^{N_t}$ symmetry instead of the single $U(1)$ of a coupled superfluid.

With tunneling, an array of tubes undergoes a dimensional crossover, or deconfinement transition, into an anisotropic three-dimensional condensate~\cite{ho2004,cazalilla2006}. Without it, the $U(1)^{N_t}$ structure is that of the sliding phases of coupled XY models and Luttinger liquids~\cite{ohern,fradkin,vishwanathcarpentier,mukhopadhyay2001}, in which interchain couplings leave each chain's phase free and modify its exponents. For dipolar bosons in a planar array, Kollath, Meyer and Giamarchi obtained the phase diagram of this sliding Luttinger liquid, with stripe and checkerboard density-wave order at strong intra-tube correlations and a collapse instability at strong dipolar coupling, and proposed the transfer of center-of-mass momentum between tubes as an experimental probe~\cite{kollath}. Momentum transfer without particle exchange is Coulomb drag in condensed matter~\cite{narozhny2016}, including between one-dimensional wires~\cite{nazarov1998}; dipolar drag has been proposed for pairs of pancake-shaped clouds~\cite{matveeva2011} and computed for bilayer Fermi gases~\cite{renklioglu2024}. Summed over an extended array, the same process is the transverse transport of longitudinal momentum, the quantity that a shear viscosity summarizes in a fluid, studied in quantum gases down to dilute Bose condensates~\cite{gust2015}.
Prior work on the dipolar tube array concerns its equilibrium phases, the inter-tube contribution to integrability breaking and thermalization~\cite{tang2018}, and, in the modeling of experiments, the self-consistent mean-field potential that the array exerts on each tube, the tubes otherwise remaining independent~\cite{zhang2026}. How longitudinal momentum spreads across many tubes in time has, to our knowledge, not been addressed.

Tube occupancies are fixed at loading and vary from tube to tube, through the envelope of the parent cloud and through fluctuations about it~\cite{tang2018,malvania2021,zhang2026}. The collective modes of the array couple through the local densities, so these variations are a quenched random potential for the transverse modes even though the Hamiltonian is uniform. Anderson localization~\cite{anderson1958} of cold atoms has used externally imposed random or quasiperiodic potentials~\cite{billy2008,roati2008,sanchezpalencia2010}, and Bogoliubov quasiparticles in such potentials have been studied along the tube axis~\cite{bilas2006,lugan2007,gurarie2008}; for bosonic excitations, stability further requires every mode energy to be positive~\cite{gurarie2003}. Here the disorder is intrinsic, it acts across the tubes, and it enters as a random density-dependent rescaling of the coupling matrix. The transverse problem is therefore closer to the harmonic chain with random masses~\cite{dyson1953,matsuda1970} than to a particle in a random potential.

We study how longitudinal momentum spreads across an array of tubes coupled by density interactions, and how this dynamics changes when the tube populations are unequal. Within zero-temperature Bogoliubov theory, we examine both the momentum correlator, termed the amplitude, and its squared magnitude, the intensity. In a uniform array the intensity spreads ballistically. Random density imbalance localizes the transverse collective modes and arrests this spreading; at fixed longitudinal momentum, the spatial decay of the time-averaged intensity measures the localization length. Equal-time correlations, by contrast, are only weakly affected by disorder and primarily reflect the interaction range. Section~\ref{sec:model} introduces the model and these observables. Sections~\ref{sec:balanced} and~\ref{sec:disorder} treat the uniform and disordered arrays, respectively. Section~\ref{sec:instability} examines the instability at strong inter-tube repulsion, and Sec.~\ref{sec:discussion} discusses experimental probes and extensions. Appendix~\ref{sec:timelengthscales} estimates the experimental time and length scales and the conditions under which tunneling can be neglected (with other Appendices devoted to fleshing out details of calculations).
\section{Models and Bogoliubov theory}
\label{sec:model}
\subsection{Model Hamiltonians}
\begin{figure}
\centering
\includegraphics[width=0.99\columnwidth]{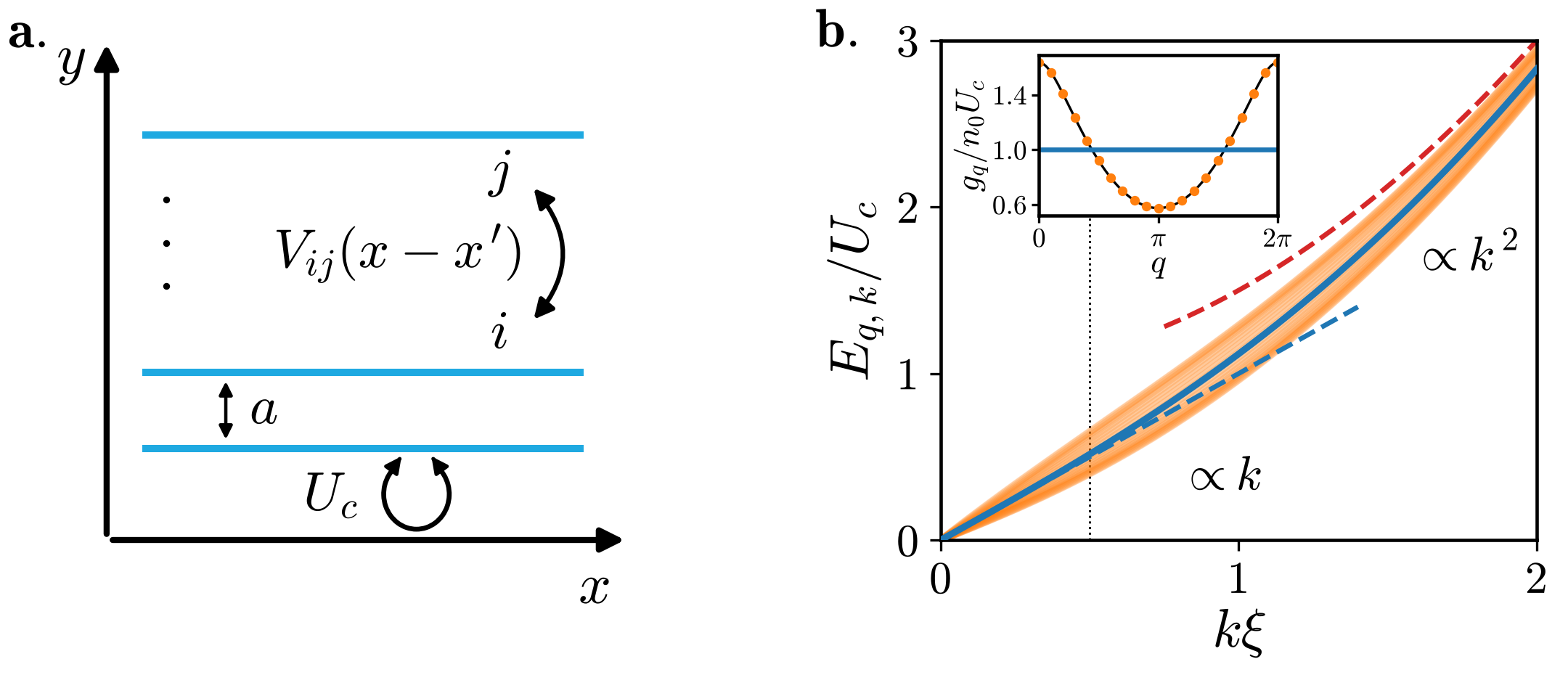}
\caption{(a) Array of $N_t$ tubes with spacing $a$, intra-tube contact interaction $U_c$ and inter-tube density coupling $V_{ij}(x-x')$. (b) Dispersion of the uniform array $E_{q,k}/U_c$ against $k\xi$ for the dipolar kernel at $V=0.3$: solid, $V=0$; shaded, the range over transverse $q$; dashed, the phonon limit $E=\sqrt{n_0U_c}\,k$ and the free-particle limit $\varepsilon_k+n_0U_c$. Inset: $g_q(k)$ across the transverse zone at $k\xi=0.5$ (dotted line), against its $V=0$ value (horizontal line).
}
\label{fig:setup}
\end{figure}

We consider a linear array of $N_t$ equidistant Bose tubes with spacing $a$ [Fig.~\ref{fig:setup}(a)], described by bosonic field operators $\Psi_i(x)$ annihilating a particle of mass $M$ at position $x$ along tube $i$, with densities $\hat n_i=\Psi_i^\dagger\Psi_i$. The Hamiltonian is
\begin{equation}
\mathcal H=\sum_i\mathcal H_i+\frac{1}{2}\sum_{i\neq j}\mathcal H_{ij},
\label{eq:model_hamiltonian}
\end{equation}
where $\mathcal H_i$ is the intratube Hamiltonian
and $\mathcal H_{ij}$ the interaction between distinct tubes,
\begin{align}
\mathcal H_i
&=\int dx\left[
\frac{\hbar^2}{2M}\partial_x\Psi_i^\dagger\partial_x\Psi_i
+\frac{U_c}{2}\hat n_i^2(x)
\right],
\label{eq:model_intratube}
\\
\mathcal H_{ij}
&=\int dx\,dx'\,
V_{ij}(x-x')\hat n_i(x)\hat n_j(x'),
\label{eq:model_intertube}
\end{align}
with $U_c>0$ the contact interaction. We consider two inter-tube potentials, nearest-neighbor (n.n.) and dipolar,
\begin{equation}
V_{ij}(x)=
\begin{cases}
V\,\delta(x)\,\delta_{|i-j|,1},
\\[2pt]
V_d\left[x^2+a^2(i-j)^2\right]^{-3/2}.
\end{cases}
\label{eq:realspace_interaction}
\end{equation}
The dipoles are polarized normal to the plane of the array, so that the dipolar coupling is repulsive and the same for every pair of tubes. The two kernels are normalized to the same nearest-neighbor bond at $k\to0$, $V\equiv V_1(0)=2V_d/a^2$ (see Eq.~\eqref{eq:dipolar_fourier}), and $V$ denotes this bond for both below.

Single-particle tunneling between tubes, $t_\perp$, is set to zero,
\begin{equation}
t_\perp=0;
\label{eq:no_tunneling}
\end{equation}
the time and momentum scales below which this is justified in a deep lattice are estimated in Appendix~\ref{sec:timelengthscales}. We use $\hbar=M=1$, measure energies in units of $n_0U_c$ with $n_0$ the mean density, lengths along the tubes in units of the healing length $\xi=1/\sqrt{n_0U_c}$, and set $a=1$ (for the realization of Appendix~\ref{sec:timelengthscales}, $a\simeq1.2\,\xi$). Longitudinal momentum is written $k$ and transverse quasimomentum $q$ throughout.

\subsection{Symmetry sectors and effective disorder}
Both models conserve the particle number $\hat N_i=\int dx\,\hat n_i(x)$ of each tube; equivalently $\mathcal H$ is invariant under $\Psi_i\to e^{i\alpha_i}\Psi_i$ for every $i$ separately. This $U(1)^{N_t}$ symmetry is the one familiar from sliding phases~\cite{ohern,fradkin,vishwanathcarpentier,kollath}. The Bogoliubov ansatz of Sec.~\ref{sec:spectrum} breaks it by picking a pattern of $\alpha_i$; nothing we compute depends on the $\alpha_i$. Translation symmetry along the tubes conserves the total longitudinal momentum of the array,
\begin{equation}
\hat P=\sum_i\hat P_i=\sum_i\int dk\,k\,a^\dagger_{i,k}a_{i,k},
\label{eq:total_momentum}
\end{equation}
where $\hat P_i$ is the momentum of tube $i$ and $a_{i,k}$ its Fourier modes. Unlike $\hat N_i$, the individual $\hat P_i$ are not conserved: momentum hops between tubes through the inter-tube interaction, and $\Pi_{ij}(t)=\langle\hat P_i(t)\hat P_j(0)\rangle_c$, Eq.~\eqref{eq:def_pi} below, is the object that describes this.

We distinguish (i) the uniform, or clean, array with $n_i=n_0$, and (ii) disordered arrays with independent tube densities,
\begin{equation}
\langle n_i\rangle=n_0,\qquad\operatorname{Var}(n_i)=\sigma^2n_0^2,
\label{eq:density_moments}
\end{equation}
drawn from the uniform distribution
\begin{equation}
n_i\in[n_0-w,n_0+w],
\qquad
w=\sqrt{3}\,\sigma n_0 .
\label{eq:halfwidth}
\end{equation}
Positivity of every density requires
\begin{equation}
\sigma<\frac{1}{\sqrt{3}}\simeq0.577 .
\label{eq:sigma_max}
\end{equation}
This represents loading fluctuations in an optical lattice. The localization length of the disordered array is written $\xiloc$ and its inverse $\gamma\equiv1/\xiloc$.

\subsection{Bogoliubov theory}
\label{sec:spectrum}
We expand $\Psi_i(x)=\sqrt{n_i}e^{i\alpha_i}+\delta\Psi_i(x)$, retain terms quadratic in the fluctuations, and write $a_{i,k}$ for the Fourier modes of $\delta\Psi_i$ along the tube. The resulting Hamiltonian is
\begin{equation}
\mathcal H=\int_{k>0}dk\,\psi_k^\dagger H(k)\psi_k,
\label{eq:quadratic_hamiltonian}
\end{equation}
where
\begin{equation}
\psi_k=
(a_{1,k},\ldots,a_{N_t,k},a_{1,-k}^\dagger,\ldots,a_{N_t,-k}^\dagger)^T
\label{eq:nambu_spinor}
\end{equation}
is the Nambu spinor and
\begin{equation}
H(k)=
\begin{pmatrix}
\hat T+\hat G & \hat G\\
\hat G & \hat T+\hat G
\end{pmatrix},
\label{eq:full-model-bogo}
\end{equation}
with $T_{ij}=\varepsilon_k\delta_{ij}$, $\varepsilon_k=k^2/2$, and the interaction matrix
\begin{equation}
G_{ij}(k)
=U_cn_i\delta_{ij}
+(1-\delta_{ij})V_{ij}(k)\sqrt{n_in_j},
\label{eq:Gmatrix}
\end{equation}
where $V_{ij}(k)=\int dx\,e^{ikx}V_{ij}(x)$ and the chemical potential of each tube, $\mu_i=U_cn_i+\sum_{j\ne i}V_{ij}(0)n_j$, has been subtracted. Density imbalance enters both the diagonal and the bonds of $\hat G$.

For dipolar interactions the partial Fourier transform is
\begin{equation}
V_{ij}(k)
=\frac{2V_d|k|}{a|i-j|}
K_1\!\left(|k|a|i-j|\right),
\label{eq:dipolar_fourier}
\end{equation}
with $K_1$ the modified Bessel function of the second kind; it reduces to $2V_d/a^2|i-j|^2$ as $k\to0$ and decays as $e^{-|k|a|i-j|}$ at large separation, so the transverse range of $\hat G(k)$ is $1/k$. The n.n. kernel has $V_{ij}(k)=V\delta_{|i-j|,1}$ with no $k$ dependence.

\label{sec:hidden}
At each $k$ we diagonalize $\hat G(k)$ with an orthogonal matrix $U_g(k)$,
\begin{equation}
U_g^T(k)\,\hat G(k)\,U_g(k)
=\operatorname{diag}\!\left[g_1(k),\ldots,g_{N_t}(k)\right],
\label{eq:G_diagonalization}
\end{equation}
so that $g_m(k)$ is the $m$-th eigenvalue of $\hat G(k)$; we omit the argument $k$ where no confusion arises. In the uniform array the eigenvectors are transverse plane waves, Eq.~\eqref{eq:g-eigen-balanced}; with disorder they are obtained numerically, once per realization and $k$. Since $\hat T=\varepsilon_k\mathbb 1$, the same rotation in particle and hole space block diagonalizes $H(k)$,
\begin{equation}
\widetilde H(k)
=\mathcal U_g^TH(k)\mathcal U_g
=\bigoplus_m
\begin{pmatrix}
\varepsilon_k+g_m & g_m\\
g_m & \varepsilon_k+g_m
\end{pmatrix},
\label{eq:def-ham-gen-tube}
\end{equation}
where $\mathcal U_g=\operatorname{diag}(U_g,U_g)$, in terms of the \emph{generalized-tube} operators
\begin{equation}
    b_{m,k}=\sum_i(U_g)_{im}a_{i,k}.
    \label{eq:gentubebasis}
\end{equation}
All mixing between physical tubes is contained in $U_g$. Each block is diagonalized by a single-mode Bogoliubov transformation,
\begin{gather}
\beta_{m,k}=u_{m,k}b_{m,k}-v_{m,k}b^\dagger_{m,-k},
\nonumber\\
u_{m,k}=\cosh\theta_{m,k},\quad v_{m,k}=\sinh\theta_{m,k},
\label{eq:coherence_factors}
\\
\cosh(2\theta_{m,k})=\frac{\varepsilon_k+g_m}{E_{m,k}},\quad E_{m,k}=\sqrt{\varepsilon_k(\varepsilon_k+2g_m)},
\label{eq:def-uk-vk-hyperbolic}
\\
\mathcal H=\int_{k>0}\!dk\sum_m E_{m,k}\big(\beta^\dagger_{m,k}\beta_{m,k}+\beta^\dagger_{m,-k}\beta_{m,-k}\big)+\text{const}.
\label{eq:diag_hamiltonian}
\end{gather}
At small $k$, $E_{m,k}\simeq|k|\sqrt{g_m}$: each transverse channel has its own longitudinal sound velocity. Because $U_g$ is orthogonal, the transverse eigenvectors are normalized and the localization of the Bogoliubov modes can be read directly from the eigenvectors of $\hat G$; the non-unitary part of the symplectic transformation is confined to the squeeze within each block.

We will need the overlap of a physical tube with two transverse modes,
\begin{align}
\mathcal M_{i,k}^{mn}
&=[U_g^Tp_iU_g]_{mn}
=(U_g)_{im}(U_g)_{in},
\nonumber\\
(p_i)_{ab}&=\delta_{ai}\delta_{bi},
\label{eq:def_M}
\end{align}
with $p_i$ the projector onto tube $i$; it is symmetric in $m,n$ and satisfies $\sum_i\mathcal M_{i,k}^{mm}=\sum_m\mathcal M_{i,k}^{mm}=1$.

\label{sec:stability}
The spectrum \eqref{eq:def-uk-vk-hyperbolic} is real when every $g_m(k)>0$. Writing $D=\operatorname{diag}(n_i)$ and
\begin{equation}
M_{ij}(k)=U_c\delta_{ij}+(1-\delta_{ij})V_{ij}(k),
\label{eq:stability_matrices}
\end{equation}
Eq.~\eqref{eq:Gmatrix} is a congruence of the density-independent matrix $M$,
\begin{equation}
\hat G(k)=D^{1/2}M(k)D^{1/2},
\label{eq:congruence}
\end{equation}
so by Sylvester's law of inertia $\hat G$ and $M$ have the same numbers of positive, negative and zero eigenvalues for any positive densities. The stability boundary is therefore independent of the density imbalance throughout the range \eqref{eq:sigma_max}; the thresholds, and what happens beyond them, are given in Sec.~\ref{sec:instability}.

\subsection{Momentum correlations}
\label{sec:correlator}
\label{sec:central}
We start by defining the average connected correlator
\begin{equation}
\Pi_{ij}(t)=\langle \hat P_i(t)\hat P_j(0)\rangle_c,
\label{eq:def_pi}
\end{equation}
evaluated in the Bogoliubov ground state. In the generalized-tube basis \eqref{eq:gentubebasis} the momentum of tube $i$ is
\begin{equation}
\hat P_i
=\int dk\,k\sum_{m,n}\mathcal M_{i,k}^{mn}
\left(
 b_{m,k}^\dagger b_{n,k}
-b_{m,-k}b_{n,-k}^\dagger
\right),
\label{eq:def-P-Mmn}
\end{equation}
and with
\begin{equation}
b_{m,k}(t)
=u_{m,k}\beta_{m,k}e^{iE_{m,k}t}
+v_{m,k}\beta_{m,-k}^\dagger e^{-iE_{m,k}t}
\label{eq:b_time_evolution}
\end{equation}
(the opposite sign convention for the time evolution replaces $\Pi_{ij}$ by its complex conjugate) the zero-temperature contractions give (Appendix~\ref{app:derivation})
\begin{equation}
\Pi_{ij}(t)
=\int dk\,k^2\sum_{m,n}
 e^{i\mathcal E_{mn,k}t}
 \mathcal O_{ij,k}^{mn}
 \sinh^2\vartheta_{mn,k},
\label{e26a}
\end{equation}
where
\begin{equation}
\mathcal E_{mn,k}=E_{m,k}+E_{n,k},
\qquad
\vartheta_{mn,k}=\theta_{m,k}-\theta_{n,k},
\label{eq:pair_variables}
\end{equation}
and
\begin{equation}
\mathcal O_{ij,k}^{mn}
=\mathcal M_{i,k}^{mn}\mathcal M_{j,k}^{nm}.
\label{eq:pair_overlap}
\end{equation}
The spatial structure is carried by the eigenvectors of $\hat G$ through $\mathcal O$, the bosonic coherence by $\sinh^2\vartheta$. Since $\hat P_i$ is bilinear, it creates a pair of quasiparticles from the vacuum: only sums of mode energies appear, and the spectrum of $\Pi$ at fixed $k$ has a threshold at $2\min_mE_{m,k}$. Each of the two mode amplitudes in $\mathcal O$ connects tube $i$ to tube $j$, which is the origin of the factor of two in Eq.~\eqref{eq:pair_factor}.

Whereas we typically expect correlation functions to be useful for characterizing important physics, $\Pi_{ij}$ suffers from two handicaps: (i) the BEC ground state is an eigenstate of the total momentum with zero fluctuation, so $\sum_j\Pi_{ij}(t)=0$ at all $t$ [Eq.~\eqref{eq:sumrule}]; (ii) disordered wave functions are random and so are the overlaps in Eq.~\eqref{eq:pair_overlap}, hence suppressing $\Pi_{ij}$ on averaging. We can circumvent these difficulties following the prescription from the theory of disorder-averaged Green functions, where the single-particle Green function is similarly compromised. To this end we define  $|\Phi_i\rangle=\hat P_i|0\rangle$ and corresponding amplitude and intensity, respectively, 
\begin{align}
\Pi_{ij}(t)&=\langle\Phi_i|e^{-i(\mathcal H-E_0)t}|\Phi_j\rangle,
\\
|\Pi_{ij}(t)|^2&=\big|\langle\Phi_i|e^{-i\mathcal Ht}|\Phi_j\rangle\big|^2.
\label{eq:transition}
\end{align}
Late time asymptotics of $|\Pi_{ij}(t)|^2$ will be used to characterize ballistic transport  next and localization in Sec.~\ref{sec:disorder}.
\section{Uniform array}
\label{sec:balanced}
For $n_i=n_0$ the interaction matrix is translationally invariant in the tube index, so its eigenvectors are Bloch waves and the channel index of Sec.~\ref{sec:hidden} acquires the meaning of a transverse momentum rather than a bare ordering label. We accordingly write $q$ in place of $m$ throughout this section. In the thermodynamic limit the eigenvalues are
\begin{equation}
g_q(k)
=n_0\left[
U_c+2\sum_{r=1}^{\infty}V_r(k)\cos(qr)
\right],
\label{eq:g-eigen-balanced}
\end{equation}
where $q$ is the transverse quasimomentum and $V_r(k)$ the interaction matrix element at separation $r$. Nearest-neighbor coupling has $V_r(k)=V\delta_{r,1}$, while the dipolar kernel with $a=1$ has
\begin{equation}
V_r(k)=\frac{2V_d|k|}{r}K_1(|k|r).
\label{eq:dipolar_Vr}
\end{equation}
The two kernels differ in how far this can be taken. For nearest neighbors the sum terminates, $g_q=n_0(U_c+2V\cos q)$, with no residual $k$ dependence. The dipolar sum has no closed form at general $k$; it converges rapidly and we evaluate it numerically, except as $k\to0$, where it reduces to the piecewise-quadratic form of Eq.~\eqref{eq:g_dipolar_smallk} and fixes the threshold of Sec.~\ref{sec:instability}. The dispersion of the uniform array is $E_{q,k}=\sqrt{\varepsilon_k[\varepsilon_k+2g_q(k)]}$.

In a complex Bloch basis the mode-pair overlap is
\begin{equation}
\mathcal O_{ij,k}^{q,q'}
=\frac{1}{N_t^2}e^{i(q'-q)(i-j)},
\label{eq:bloch_overlap}
\end{equation}
so that the transverse Fourier transform of the correlator reads ($q_\pm=q\pm\frac{Q}{2}$)
\begin{equation}
\Pi_Q(t)
=\int dq\,dk\,k^2
 e^{i\mathcal E_{q_-,q_+,k}t}
 \sinh^2\vartheta_{q_-,q_+,k}.
\label{eq:PiQ_clean}
\end{equation}

\subsection{Amplitude}
\label{sec:amplitude}
\subsubsection{Kinematics}
\label{sec:kinematics}
\begin{figure}
\centering
\includegraphics[width=\linewidth]{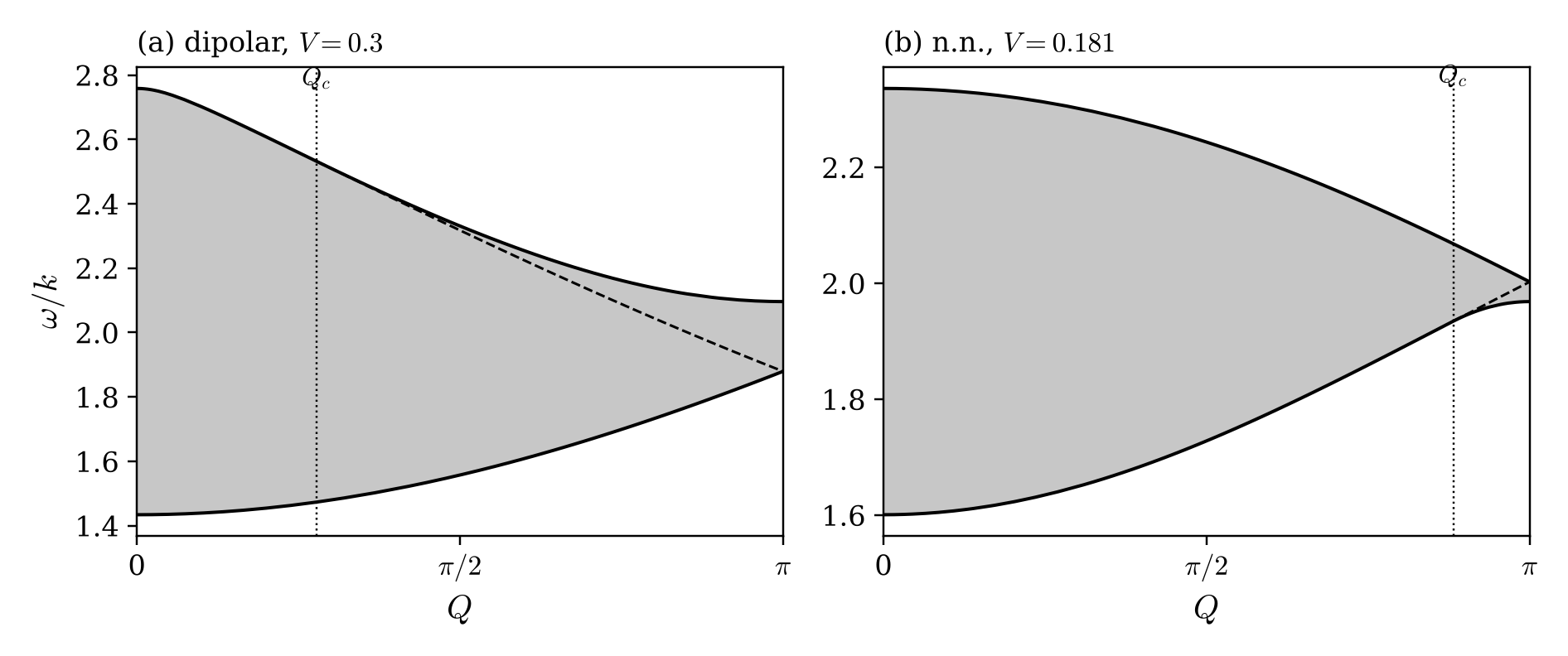}
\caption{Support of the pair spectrum $\Pi_Q(\omega|k)$, Eq.~\eqref{eq:PiQomega}, at $k=0.1$ (grey), for (a) the dipolar kernel at $V=0.3$ and (b) the n.n. kernel at the matched bond $V=0.181$. Solid lines are the band edges, critical points of $\Omega_Q(q)$ at $q=0$ and $q=\pi$ or at the pair $\pm q_c$; the dashed line is the interior critical point that exists for $Q>Q_c$ (dotted). Every line is an inverse-square-root singularity; the weight $S_Q$, not shown, vanishes as $Q^2$ at small $Q$ and is largest near $Q=\pi$. For (b), Eq.~\eqref{eq:Omega_nn} gives $Q_c=\pi-2V=2.78$ and the support $[2-V^2,2]$ at $Q=\pi$, against $2.77$ and $[1.968,2.003]$ from the full dispersion.}
\label{fig:pair_continuum}
\end{figure}

At fixed $k$ the amplitude is a sum over pairs of transverse modes, and its frequency content is the weighted joint density of states of the pair continuum,
\begin{equation}
\Pi_Q(\omega|k)=\sum_q S_Q(q)\,\delta\big(\omega-\Omega_Q(q)\big),
\label{eq:PiQomega}
\end{equation}
with $\Omega_Q(q)=E_{q_-,k}+E_{q_+,k}$, $S_Q(q)=\sinh^2(\theta_{q_-,k}-\theta_{q_+,k})$ and $\Pi_Q(k,t)=\int d\omega\,e^{i\omega t}\Pi_Q(\omega|k)$, the $k^2$ and normalization factors being those of Eq.~\eqref{eq:PiQ_clean}. In the variables $q_\pm$ both $\Omega_Q$ and $S_Q$ are even in $q$, so $q=0$ and $q=\pi$ are critical points of $\Omega_Q$ at every $Q$ and all other critical points come in pairs $\pm q_c$. The support of Eq.~\eqref{eq:PiQomega} is the range of $\Omega_Q(q)$, Fig.~\ref{fig:pair_continuum}: the doubled band $[2E_{\min},2E_{\max}]$ at $Q=0$, where $S_Q\propto Q^2$, narrowing toward $Q=\pi$. Each critical point is a van Hove singularity, $\Pi_Q(\omega|k)\propto S_Q(q_c)\,|\Omega_Q''(q_c)|^{-1/2}|\omega-\Omega_Q(q_c)|^{-1/2}$, one-sided at the edges and two-sided inside the band.

For the nearest-neighbor kernel in the phonon limit, $E_{q,k}\to k\sqrt{g_q}$ with $g_q=1+2V\cos q$, the structure is explicit to second order in $V$:
\begin{equation}
\frac{\Omega_Q(q)}{k}=2+2V\cos\tfrac Q2\cos q-\tfrac{V^2}{2}\big[1+\cos Q\cos2q\big].
\label{eq:Omega_nn}
\end{equation}
The point $q=0$ is the maximum for all $Q$, the upper edge $2+2V\cos\frac Q2-V^2\cos^2\frac Q2$. The point $q=\pi$ is the minimum for $Q<Q_c$, with $Q_c=\pi-2V$ to this order; beyond $Q_c$ a pair of minima splits off at $q=\pi\pm\delta$, $\cos\delta=\cos\frac Q2/(V|\cos Q|)$, and $q=\pi$ becomes a local maximum inside the band (dashed line in Fig.~\ref{fig:pair_continuum}), where the singularity is two-sided; at $Q_c$ itself the curvature at $q=\pi$ vanishes and the onset is $|\omega-\omega_c|^{-3/4}$. The band narrows from width $4V$ at $Q=0$ to $V^2$ at $Q=\pi$, where $\delta=\pi/2$ and the support is $[2-V^2,2]$: the first-order terms cancel for every $q$ because $g_{q+\pi}=2-g_q$ for a bipartite kernel. The dipolar kernel has the same critical-point structure, with $Q_c$ further from $\pi$ ($Q_c=0.87$ at $k=0.1$) and a band of width $O(V)$ at $Q=\pi$, since the even-separation bonds do not cancel; its interior line crosses the band from the upper edge at $Q_c$ to the lower edge at $Q=\pi$. Two consequences run through the rest of the section: there is no weight below $2\min_qE_{q,k}$ at any $Q$, and the staggered pair $Q=\pi$ is nearly dispersionless, so it dephases last.

\subsubsection{Equal time, spatial decay profiles}
\label{sec:realspace}

The squeeze factor in Eq.~\eqref{eq:PiQ_clean} separates into functions of the two transverse momenta,
\begin{equation}
\sinh^2(\theta_q-\theta_{q'})
=s_qc_{q'}+c_qs_{q'}-2h_qh_{q'},
\label{eq:squeeze_factorize}
\end{equation}
where
\begin{equation}
s_q=\frac{\varepsilon_k+g_q}{2E_{q,k}}-\frac{1}{2},
\qquad
c_q=\frac{\varepsilon_k+g_q}{2E_{q,k}}+\frac{1}{2},
\label{eq:sc_definitions}
\end{equation}
and
\begin{equation}
h_q=\frac{g_q}{2E_{q,k}}.
\label{eq:h_definition}
\end{equation}
Equation~\eqref{eq:squeeze_factorize} is the hyperbolic angle-addition identity, and it factorizes because the two mode labels enter $\vartheta_{qq'}$ only through their own squeeze angles. This is what allows the double transverse-mode sum to collapse. For $x_q\in\{s_q,c_q,h_q\}$ define the Fourier coefficients
\begin{equation}
X_r(k,t)
=\int\frac{dq}{2\pi}
 e^{iE_{q,k}t-iqr}x_q,
\label{eq:Xr_definition}
\end{equation}
with $r=i-j$ the transverse separation. All three integrands are even in $q$, so $X_{-r}=X_r$, and
\begin{equation}
\Pi_r(t)
=2\int dk\,k^2
\left[S_r(k,t)C_r(k,t)-H_r^2(k,t)\right].
\label{eq:clean_realspace}
\end{equation}
This expression is exact within the quadratic theory at all $r$ and $t$.

The coherence factors satisfy the pointwise identity
\begin{align}
c_qs_q-h_q^2
&=\frac{(\varepsilon_k+g_q)^2-g_q^2}{4E_{q,k}^2}-\frac{1}{4}
=0,
\label{eq:cs_identity}
\end{align}
which holds by the definition of $E_{q,k}$. Applying Parseval's theorem to Eq.~\eqref{eq:clean_realspace} then gives
\begin{equation}
\sum_j\Pi_{ij}(t)=0,
\label{eq:sumrule}
\end{equation}
the correlator form of the conservation law in Eq.~\eqref{eq:total_momentum}, and a dynamical statement rather than an equal-time one. Since $\Pi_{ii}(t)\neq0$, Eq.~\eqref{eq:sumrule} forces a zero crossing in $|i-j|$: the profile cannot be a simple positive exponential at any time, which matters for the fitting procedure of Sec.~\ref{sec:disorder-quench}.

\begin{figure*}
\centering
\includegraphics[width=1.9\columnwidth]{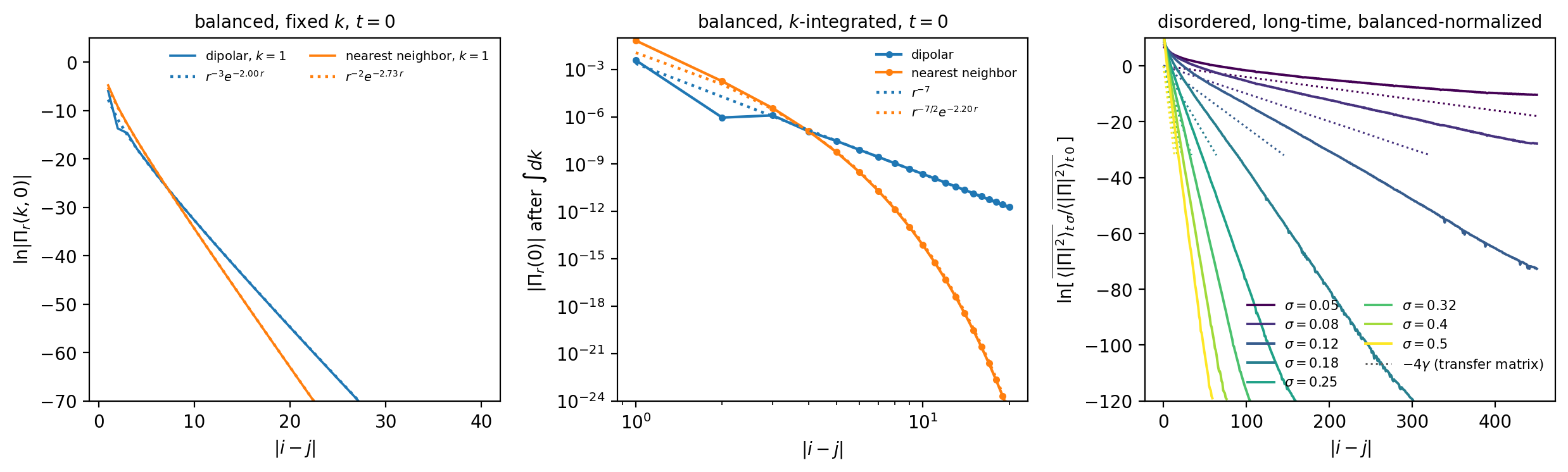}
\caption{Spatial structure of the momentum correlator at $V=0.3$. Panel (a) shows the fixed-$k$ profile of the uniform array at $k=1$ for both kernels, from Eq.~\eqref{eq:clean_realspace}. Dotted lines give the predicted asymptotic forms of Eqs.~\eqref{eq:xi_clean_NN} and \eqref{eq:xi_clean_dip}, anchored to the data at $r=8$: $r^{-2}e^{-2\kappa_Er}$ with $2\kappa_E=2.73$ for nearest neighbors, and $r^{-3}e^{-2|k|r}$ for dipolar. Both the decay rates and the algebraic prefactors are reproduced. Panel (b) shows the same profiles after integration over longitudinal momentum, on log-log axes, against the $r^{-7}$ law of Eq.~\eqref{eq:clean_kint_dip} and the $r^{-7/2}e^{-2r\operatorname{arccosh}(U_c/2V)}$ form of Eq.~\eqref{eq:clean_kint_NN}. Panel (c) shows the normalized long-time average of Eq.~\eqref{eq:clean_normalized} for disordered dipolar arrays at $N_t=1000$ and $k=1$, whose slope is the localization rate $4\gamma$ of Eq.~\eqref{eq:correlator_gamma}. Dotted lines give that slope with $\gamma$ from the transfer matrix of Appendix~\ref{app:num_tm}, nothing being fitted to these curves; the measured decay approaches it from above as $N_t/\xiloc$ grows, the shortfall at weak disorder being the finite-ring bias quantified in Fig.~\ref{fig:xi_comp}. Curves are truncated at the roundoff floor, and the rates quoted in Sec.~\ref{sec:disorder-quench} are fitted over $d/\xiloc\in[3,8]$.}

\label{fig:clean_quench}
\end{figure*}

At fixed longitudinal momentum the decay is governed by the branch point of the integrand of Eq.~\eqref{eq:Xr_definition} lying nearest the real $q$ axis: one at $|\operatorname{Im}q|=\kappa$ makes each Fourier coefficient decay at $\kappa$, and $\Pi_r$, being bilinear in them, at $2\kappa$. The two kernels place it differently. For nearest-neighbor coupling $\hat g(q)$ is entire, so the branch point of $E_{q,k}$ controls and $\kappa_E(k)=\operatorname{arccosh}z_k$ with $z_k=(k^2/2+2n_0U_c)/(4n_0V)$. For the dipolar kernel the Bessel sum converges only inside $|\operatorname{Im}q|<|k|$, and for the parameters used here that strip edge is nearer, so the rate is $|k|$ itself. Hence
\begin{align}
\Pi^{\rm NN}_r(k,0)&\sim C_{\rm NN}(k)\,r^{-2}e^{-2\kappa_E(k)r},
\label{eq:xi_clean_NN}
\\
\Pi^{\rm dip}_r(k,0)&\sim C_{\rm dip}(k)\,r^{-3}e^{-2|k|r},
\label{eq:xi_clean_dip}
\end{align}
with $C(k)$ smooth. The two exponents differ because the correlator factorizes into transforms carrying reciprocal powers of $E_{q,k}$, whose singular behavior coincides for the dipolar kernel and cancels for the nearest-neighbor one; the derivation is given in Appendix~\ref{app:clean_factorized}. Both forms are confirmed against Eq.~\eqref{eq:clean_realspace} in Fig.~\ref{fig:clean_quench}, prefactors included. The dipolar correlator therefore carries an interaction-induced exponential envelope even though its transverse eigenmodes are Bloch waves.

The longitudinal-momentum integral treats the two kernels differently, and the difference is qualitative. For nearest-neighbor coupling $\kappa_E$ is minimized at $k=0$ with quadratic curvature, $\kappa_E(k)=\kappa_0+ak^2+O(k^4)$, where
\begin{equation}
a=\frac{1}{4n_0V\sqrt{(U_c/2V)^2-1}}
\label{eq:kappa_curvature}
\end{equation}
sets the width in $k$ that contributes at large $r$. The amplitude vanishes quadratically as well, since the explicit $k^2$ in Eq.~\eqref{e26a} survives the cancellation of the $1/k$ divergences between $S_rC_r$ and $H_r^2$. Laplace integration then contributes a further $r^{-3/2}$, and
\begin{equation}
\Pi_r(0)
\sim \frac{C}{r^{7/2}}
\exp\!\left[-2r\operatorname{arccosh}\!\left(\frac{U_c}{2V}\right)\right].
\label{eq:clean_kint_NN}
\end{equation}
For dipolar interactions the fixed-$k$ rate in Eq.~\eqref{eq:xi_clean_dip} instead \emph{vanishes} linearly as $k\to0$, so no exponential survives the integration and
\begin{equation}
\Pi_r(0)\sim \frac{C}{r^7}.
\label{eq:clean_kint_dip}
\end{equation}
The dipolar profile of the uniform array is thus a pure power law. Both regimes are shown in Fig.~\ref{fig:clean_quench}. This power-law background does not enter the localization analysis of Sec.~\ref{sec:disorder}, which works at fixed $k$ and with the long-time average, whose profile in the uniform array is flat (Sec.~\ref{sec:disorder-quench}).

\subsubsection{Late time asymptotics, $Q\to 0$}
\label{sec:viscosity}
Expanding Eq.~\eqref{eq:PiQ_clean} to order $Q^2$,
\begin{align}
\Pi_Q(t)
&=Q^2A_\Lambda(t)+O(Q^4),
\nonumber\\
A_\Lambda(t)
&=\frac{1}{4}\int_0^\Lambda dk\,k^2
\int_0^{2\pi}dq\,
\frac{(g'_q)^2}{(\varepsilon_k+2g_q)^2}
 e^{2iE_{q,k}t},
\label{eq:A_explicit}
\end{align}
where $g'_q=\partial_qg_q$ and $\Lambda$ is a longitudinal ultraviolet cutoff. We define $\eta=A_\Lambda(0)$ for later use, see also Appendix~\ref{app:numerics} for details of setting the cutoff. 

\begin{figure}
\centering
\includegraphics[width=\linewidth]{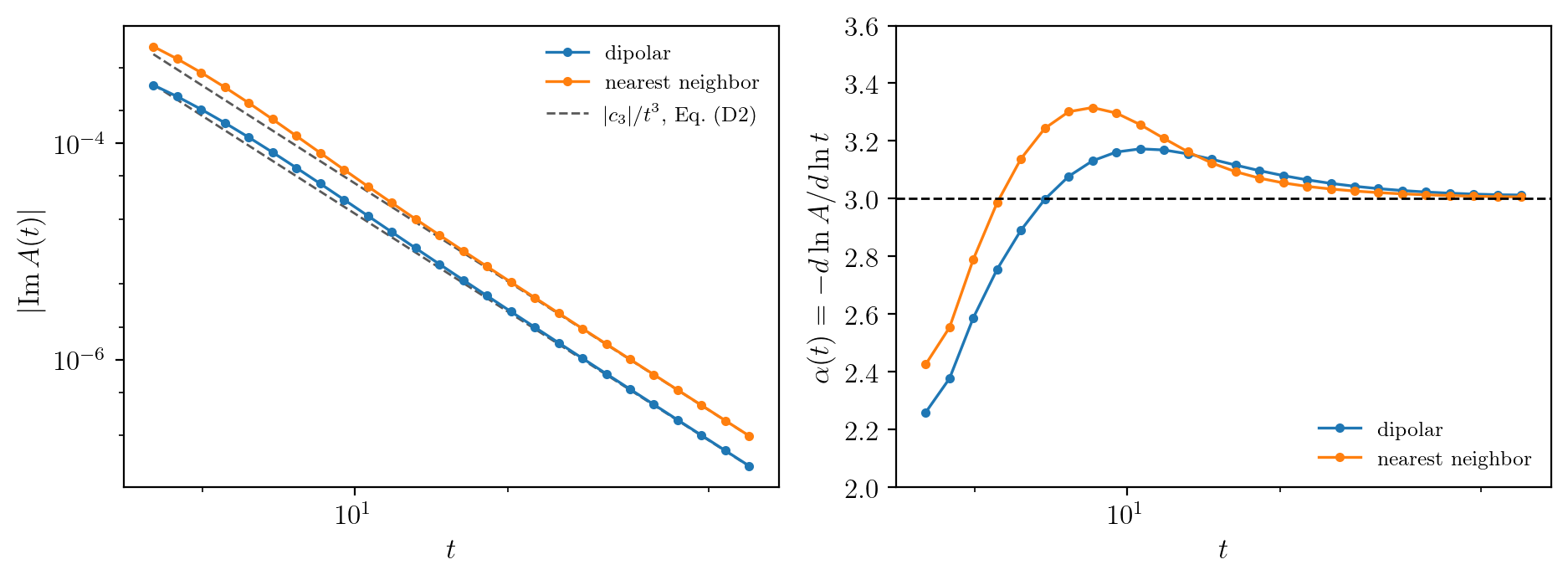}
\caption{Temporal decay of the small-$Q$ response of the uniform array at $V=0.3$ for both kernels. The left panel shows $|{\rm Im}\,A(t)|$ against the $t^{-3}$ law of Eq.~\eqref{eq:A_powerlaw}, with $c_3$ evaluated from Eq.~\eqref{eq:ImA_asymptotic}. The nearest-neighbor curve uses the closed form of Eq.~\eqref{eq:A_master}; the dipolar curve is Eq.~\eqref{eq:A_explicit} by direct quadrature, that reduction not applying when the band disperses with $k$. The right panel shows the local exponent $\alpha(t)=-d\ln A/d\ln t$, which overshoots $3$ before settling onto it once the subleading $t^{-5}$ correction has died off. The cutoff used for it is discussed in Appendix~\ref{app:num_cutoff}.}
\label{fig:clean_t3}
\end{figure}

The long-time behavior follows from the $k\to0$ endpoint of Eq.~\eqref{eq:A_explicit} and holds for either kernel. There $E_{q,k}\to k\sqrt{g_q(0)}$, so the phase is linear in $k$ with no stationary point, while the amplitude vanishes as $k^2$; the endpoint expansion of Appendix~\ref{app:J_asymptotics} then gives
\begin{equation}
A(t)\sim \frac{c_3}{t^3},
\qquad
t\rightarrow\infty,
\label{eq:A_powerlaw}
\end{equation}
with $c_3$ fixed by the transverse interaction through $\int dq\,(g'_q)^2/g_q^{7/2}$. The $t^{-3}$ decay makes $\int_0^\infty dt\,A(t)$ finite; it is a sum rule over the pair spectrum, and the next paragraph explains why it is no transport coefficient. The approach to the power law is shown in Fig.~\ref{fig:clean_t3}.

Equations~\eqref{eq:A_explicit} and \eqref{eq:A_powerlaw} should be read against the hydrodynamics of a conserved density. There the transverse Fourier component of the density relaxes as $\chi_Qe^{-DQ^2t}$: the static susceptibility $\chi_Q$ tends to a constant as $Q\to0$, the time dependence enters only through $Q^2t$, and the rate vanishes with $Q$ because the total is conserved. The total longitudinal momentum $\sum_i\hat P_i$ is conserved here as well, but the Bogoliubov ground state is an eigenstate of it with zero fluctuation, so $\Pi_{Q=0}(t)$ vanishes identically, and the small-$Q$ correlator is $Q^2A(t)$ with the $Q$ and $t$ dependence factorized. What decays in $A(t)$ is the phase coherence of the two-quasiparticle continuum of Eq.~\eqref{eq:PiQ_clean}: $\hat P_i$ creates a pair with energy $E_{q,k}+E_{q+Q,k}$, the spectrum of $\Pi$ has a threshold at $2\min_qE_{q,k}$ and vanishes as $\omega^2$ after the $k$ integration, and the $t^{-3}$ law is the $k\to0$ endpoint of that continuum. Nothing diffusive appears at any $Q$ because the quadratic theory has no quasiparticle collisions. A momentum disturbance on one tube crosses the array as a wave packet with the transverse group velocity $\partial_qE_{q,k}$, and the width of the momentum profile grows ballistically until the ring closes (Sec.~\ref{sec:no_transport}). A viscosity, the $\omega\to0$ limit of a stress correlator, would require the intraband channel $E_m-E_n$, which is empty at $T=0$ and opens with thermal quasiparticles, together with the collisions that make it dissipative. We therefore speak of transverse momentum transport throughout, and of $\eta$ as the equal-time small-$Q$ coefficient: it measures how far the momentum profile extends at $t=0$, which is set by the interaction range, and Sec.~\ref{sec:eta} finds it accordingly insensitive to localization.

For nearest-neighbor coupling the same result is available in closed form. Since $g_q$ is then independent of $k$, the change of variables $k=2\sqrt{g_q}\,x$ reduces Eq.~\eqref{eq:A_explicit} to
\begin{equation}
A_\Lambda(t)
=\frac{1}{2}\int_0^{2\pi}dq\,
\frac{(g'_q)^2}{\sqrt{g_q}}
J\!\left(4g_qt,\frac{\Lambda}{2\sqrt{g_q}}\right),
\label{eq:A_master}
\end{equation}
with the master integral
\begin{equation}
J(s,X)
=\int_0^Xdx\,
\frac{x^2}{(1+x^2)^2}
 e^{isx\sqrt{1+x^2}},
\label{eq:J_def}
\end{equation}
which is ultraviolet finite at $t=0$ and reproduces Eq.~\eqref{eq:A_powerlaw} with the same coefficient. Appendix~\ref{app:dipolar_integral} collects the transverse integrals. At the instability threshold the logarithmic divergence of the nearest-neighbor one is cancelled by a vanishing prefactor, so the singular behavior of $\eta$ there originates in the longitudinal integral instead. The long-time coefficient behaves differently, since $\int dq\,(g'_q)^2/g_q^{7/2}$ carries a higher power of $g_q$ and does diverge as $g_\pi\to0$.

\subsection{Intensity}
\label{sec:spreading}
\begin{figure}
\centering
\includegraphics[width=\linewidth]{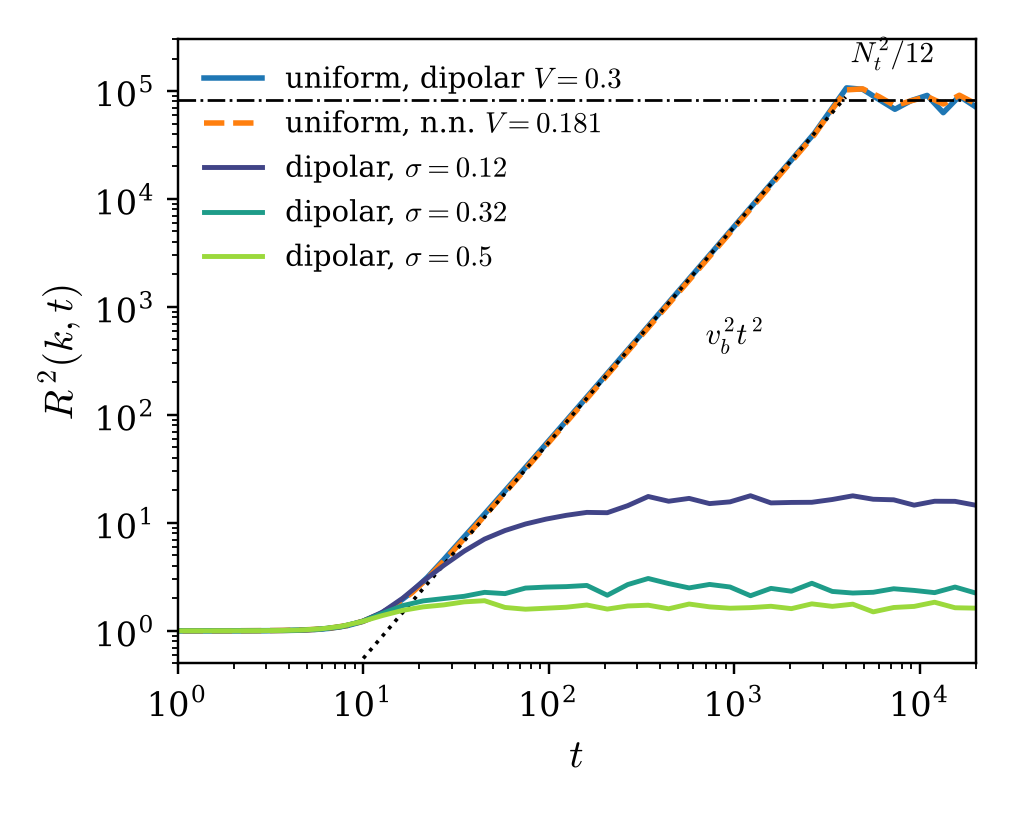}
\caption{Spatial variance $R^2(k,t)$ of the intensity profile, Eq.~\eqref{eq:R2_general}, at $k=1$, $N_t=1000$: uniform dipolar ($V=0.3$) and n.n. ($V=0.181$) arrays, the ballistic law $v_b^2t^2$ of Eq.~\eqref{eq:vb} (dotted), the flat-ring value $N_t^2/12$ (dash-dotted), and disordered dipolar arrays at three $\sigma$ (4 realizations, 20 source tubes each), which leave the ballistic line at $t\sim\xiloc/v_b$ and saturate at the values of Fig.~\ref{fig:momentum_spread}.}
\label{fig:R2_time}
\end{figure}

The effective mean-square distance the excitations have moved via ($d_{ij}=\rm{min}(|i-j|, N_t-|i-j|)$)
\begin{equation}
R^2(k,t)\equiv\frac{\sum_{j\neq i}d_{ij}^2\,|\Pi_{ij}(k,t)|^2}{\sum_{j\neq i}|\Pi_{ij}(k,t)|^2}.
\label{eq:R2_general}
\end{equation}
It is relatively straightforward to see that this quantity is not dominated by small $Q$ regime where $\Pi_Q$ becomes suppressed
\begin{align}
\sum_r|\Pi_r(k,t)|^2&=\frac1{N_t}\sum_Q|\Pi_Q(k,t)|^2,
\nonumber\\
\sum_rr^2|\Pi_r(k,t)|^2&=\frac1{N_t}\sum_Q|\partial_Q\Pi_Q(k,t)|^2
\label{eq:R2_parseval}
\end{align}
The denominator is the probability that the pair created on tube $i$ is found, at time $t$, as a pair on a single tube. Written in real space, $\Pi_r(k,t)=N_t^{-2}\sum_{q,q'}S(q,q')\,e^{i(q-q')r+i(E_{q,k}+E_{q',k})t}$ with $S=\sinh^2(\theta_q-\theta_{q'})$, and for $1\ll t\ll N_t/v$ the double sum is dominated by the stationary points $v_q=-v_{q'}=r/t$, $v_q=\partial_qE_{q,k}$: the two quasiparticles must have moved to the same tube, so only pairs with equal transverse velocity contribute. These are the critical points of $\Omega_Q(q)$, $\partial_q\Omega_Q=0$, i.e. the van Hove lines of Fig.~\ref{fig:pair_continuum}, which for the pairs that carry the signal lie at $Q$ between $Q_c$ and $\pi$. Each such pair gives $|\Pi_r|^2\propto t^{-2}$, uniformly in $r$ inside the cone $|r|<v_{\max}t$, so the denominator decays as $1/t$ while the numerator, with $r^2\propto t^2$, grows as $t$. The spread is ballistic,
$R^2(k,t)=v_b^2t^2$ with
\begin{equation}
v_b^2=\frac{\displaystyle\int dq\,dq'\,v_q^2\,S(q,q')^2\,\delta(v_q+v_{q'})}{\displaystyle\int dq\,dq'\,S(q,q')^2\,\delta(v_q+v_{q'})},
\label{eq:vb}
\end{equation}
the mean-square group velocity of the co-moving pairs, weighted by their creation amplitude. At $k=1$ Eq.~\eqref{eq:vb} gives $v_b^2=5.55\times10^{-3}$ for the dipolar kernel at $V=0.3$ and $5.38\times10^{-3}$ for nearest neighbors at the matched bond, against $v_{\max}=0.162$ and $0.164$; the profile is roughly uniform inside the cone with a sharp front at $v_{\max}t$, and Fig.~\ref{fig:R2_time} shows $R^2(k,t)$ on the ballistic line over two decades in $t$ before the ring closes at $t\simeq N_t/2v_{\max}$ and $R^2$ settles on the flat-ring value $N_t^2/12$. At $t\lesssim10$ the profile is still the equal-time one, $R^2\simeq1$, and the first departure is quadratic in $t$ with a smaller coefficient fixed by the energy moments $\sum_q(E_{q,k}+E_{q+Q,k})^nS$, $n\le2$, which we have checked against the numerics but do not need. Nothing in Eq.~\eqref{eq:vb} is dissipative: $v_b$ is set by the Bogoliubov dispersion alone, and the only decay in the uniform array, that of the denominator, is the separation of the two quasiparticles of a pair. With disorder the same $R^2(k,t)$ follows the ballistic line until $t\sim\xiloc/v_b$ and then saturates (Fig.~\ref{fig:R2_time}, Sec.~\ref{sec:no_transport}).

\section{Disordered arrays}
\label{sec:disorder}

Density imbalance turns $\hat G(k)$ into a one-dimensional disordered matrix in the tube index, with the diagonal $U_cn_i$ and the bonds $V_{ij}(k)\sqrt{n_in_j}$ both random. Each $k$ is a separate localization problem, and for the dipolar kernel the range of $\hat G(k)$ is $1/k$, so the eigenvectors cannot decay faster than the bonds: the localization rate is bounded, $\gamma(k)\le|k|$, and at $k\to0$, where the bonds fall as $1/r^2$, localization is at most algebraic. The analysis below is at fixed $k$, mostly $k=1$, where $\gamma$ is well inside this bound except at the strongest disorder (Fig.~\ref{fig:xi_comp}); the $k$ integration is performed only where the uniform-array asymptotics require it. We write $\Pi_{ij}(k,t)$ and $\Pi_Q(k,t)$ for the correlator at fixed $k$, and $\Pi_{ij}(t)$, $\Pi_Q(t)$ for its $k$ integral; in this section an unadorned $\Pi_{ij}$ inside a long-time average is the fixed-$k$ object.

\subsection{Localization of the collective modes}
\label{sec:localization}
\label{sec:ipr}

\begin{figure*}
\centering
\includegraphics[width=2\columnwidth]{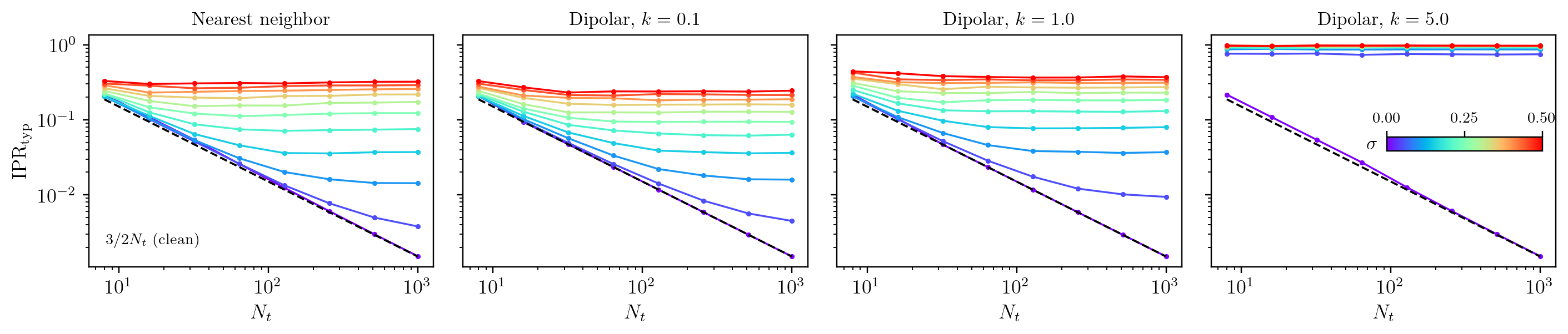}
\caption{Typical IPR, Eq.~\eqref{eq:ipr}, of band-center modes against $N_t$ at $V=0.3$, for the n.n. kernel and the dipolar kernel at three $k$; color gives $\sigma$, 100 realizations per point. Dashed: the clean value $3/2N_t$.
}
\label{fig:ipr_dist}
\end{figure*}

We characterize the transverse extent of mode $m$ by its inverse participation ratio (IPR),
\begin{equation}
\mathcal I_m
=\sum_{i=1}^{N_t}
\left(\mathcal M_{i,k}^{mm}\right)^2
=\sum_{i=1}^{N_t}(U_g)_{im}^4,
\label{eq:ipr}
\end{equation}
a property of the eigenvectors of $\hat G$ alone. In the uniform array real standing waves give $\mathcal I_m=3/2N_t$ (a complex plane wave would give $1/N_t$); Fig.~\ref{fig:ipr_dist} shows $\mathcal I$ moving off this floor to an $N_t$-independent value as $\sigma$ increases, for both kernels and several $k$.

\begin{figure}
\centering
\includegraphics[width=\linewidth]{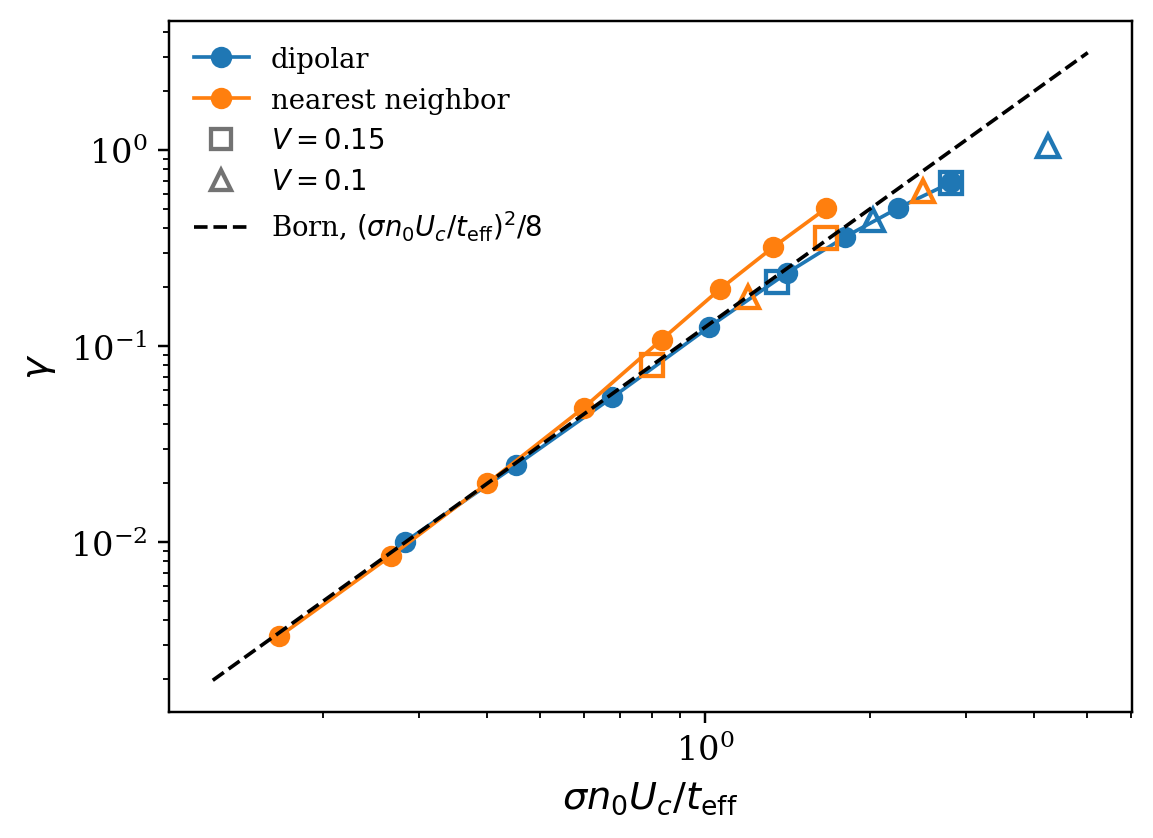}
\caption{Localization rate at the band center against $\sigma n_0U_c/t_{\rm eff}$, $k=1$. Filled circles: disorder sweep at $V=0.3$; open symbols: coupling scan at $V=0.15$ and $0.1$. Dashed: Eq.~\eqref{eq:gamma_born}.}
\label{fig:V_dependence}
\end{figure}

The asymptotic localization rate $\gamma=1/\xiloc$ is obtained from the transfer matrix of $\hat G(k)$ (Appendix~\ref{app:num_tm}). It depends on where the mode sits in the band, and $\gamma$ denotes its minimum over the band: the most extended modes dominate the long-distance tail of the correlator, so that is the rate the correlator measures. At weak disorder the rate at the band center $g=n_0U_c$ is given by the Born approximation,
\begin{equation}
\gamma_{\rm Born} = \frac{\sigma^2 n_0^2 U_c^2}{8 t_{\rm eff}^2}, \qquad t_{\rm eff} = \Big|\sum_r r\,t_r\sin(q_cr)\Big|,
\label{eq:gamma_born}
\end{equation}
where $t_r=n_0V_r(k)$ is the uniform-array bond at separation $r$ and $q_c$ the band-center quasimomentum, $g_{q_c}=n_0U_c$, so that $2t_{\rm eff}=|\partial_qg_q|_{q_c}$ is the band-center group velocity and Eq.~\eqref{eq:gamma_born} is Eq.~\eqref{eq:gamma_random_mass} evaluated there (Appendix~\ref{app:num_tm}); for nearest neighbors $q_c=\pi/2$ and $t_{\rm eff}=t_1$.
Since $t_{\rm eff}\propto V$, $\gamma$ depends on $\sigma$ and $V$ only through $\sigma n_0U_c/t_{\rm eff}$ in this regime; Fig.~\ref{fig:V_dependence} shows the disorder sweep, the coupling scan and both kernels collapsing onto Eq.~\eqref{eq:gamma_born}. Beyond the Born regime the kernels part: the dipolar rate falls below the line, the n.n. rate rises above it, and in both cases $\sigma$ and $V$ no longer enter through one combination, because the effective random potential $g(1/n_i-1/n_0)$ is not Gaussian and carries no factor of $V$ at second order (Appendix~\ref{app:num_tm}). Here and in all localization figures the n.n. coupling is matched to the dipolar first-neighbor bond at the $k$ used, $V_{\rm nn}=VkK_1(k)=0.181$ at $k=1$, so that the two arrays are compared at equal bond and comparable $N_t/\xiloc$.

The IPR follows the disorder dependence of $\gamma$, reproducing its power law and magnitude, but $2\mathcal I_{\rm typ}/\gamma$ drifts across the disorder range (Fig.~\ref{fig:xi_comp}): the fourth moment in Eq.~\eqref{eq:ipr} weights the log-normal amplitude fluctuations riding on the envelope (Appendix~\ref{app:num_env}), and $\mathcal I_{\rm typ}$ averages over the band whereas $\gamma$ is its minimum. The drift is systematic, so the IPR detects localization but does not measure $\gamma$; we quote the transfer-matrix rate throughout.

\subsection{Long-time momentum correlator}
\label{sec:disorder-quench}

For one disorder realization the fixed-$k$ correlator is
\begin{equation}
\Pi_{ij}(k,t)
=\sum_{m,n}c_{ij}^{mn}e^{i\mathcal E_{mn,k}t},
\label{eq:pi_mode_expansion}
\end{equation}
with $c_{ij}^{mn}$ the overlap and coherence factors of Eq.~\eqref{e26a}, and for a nondegenerate spectrum its infinite-time average is
\begin{equation}
\left\langle|\Pi_{ij}|^2\right\rangle_t
=2\sum_{m,n}|c_{ij}^{mn}|^2
\label{eq:timeavg}
\end{equation}
(the factor two from $c^{mn}=c^{nm}$; $c^{mm}=0$), evaluated in closed form without a time grid (Appendix~\ref{app:num_pi}). A nonlinear functional of the time record is needed: at any fixed $t$ the correlator is a smooth matrix function of $\hat G$, Eq.~\eqref{eq:PM_form}, whose decay in $|i-j|$ is set by the range of the kernel and the spectrum of $\hat G$, not by the localization of its eigenvectors. The time average kills all cross terms and leaves a sum of squared mode amplitudes, which does decay at the localization rate. A localized eigenvector has envelope $e^{-\gamma r}$; the pair overlap \eqref{eq:pair_overlap} contains two amplitudes connecting $i$ to $j$, so
\begin{equation}
|\Pi_{ij}|\sim e^{-2\gamma|i-j|}
\label{eq:pair_factor}
\end{equation}
and
\begin{equation}
\left\langle|\Pi_{ij}|^2\right\rangle_t
\sim e^{-4\gamma|i-j|}.
\label{eq:correlator_gamma}
\end{equation}
We average $\ln\langle|\Pi_{ij}|^2\rangle_t$ over realizations and over tube pairs at fixed separation (the annealed average is dominated by rare slowly decaying realizations and does not converge with the fit window), and plot
\begin{equation}
\ln\!\left[
\frac{
\overline{\left\langle|\Pi_{ij}|^2\right\rangle_t}_{\sigma}
}{
\overline{\left\langle|\Pi_{ij}|^2\right\rangle_t}_{0}
}
\right],
\label{eq:clean_normalized}
\end{equation}
where the uniform-array denominator is independent of $|i-j|$ beyond a few sites (the dephased profile of the uniform array is flat on the ring, Sec.~\ref{sec:no_transport}) and only fixes the overall scale.
The rate is fitted over $3\le d/\xiloc\le8$ (Appendix~\ref{app:numerics}).

\begin{figure}
\centering
\includegraphics[width=\linewidth]{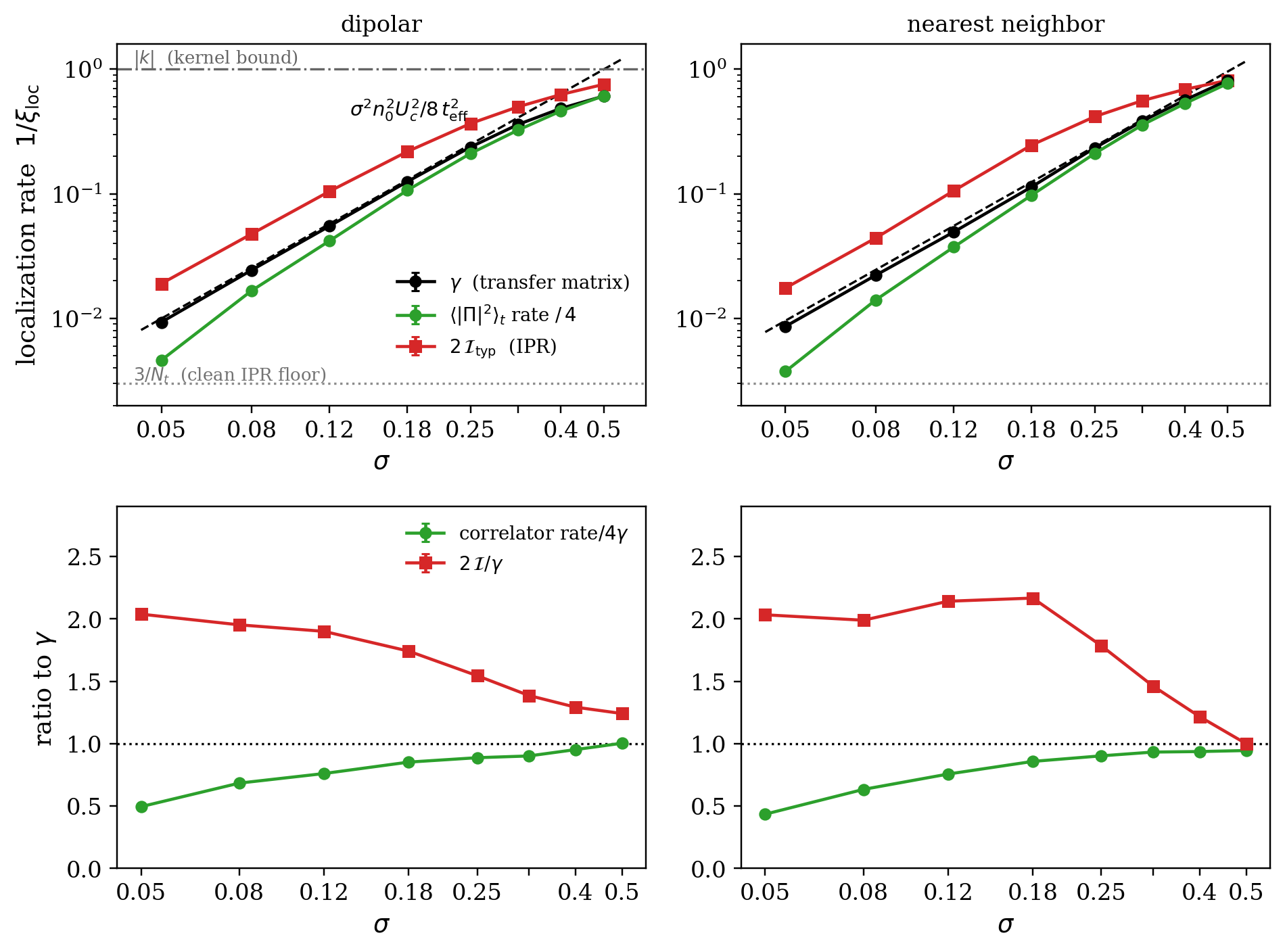}
\caption{Localization rate from three diagnostics at $k=1$, $N_t=1000$, 100 realizations; dipolar at $V=0.3$ (left), n.n. at the matched bond $0.181$ (right). Top: transfer-matrix $\gamma$ (black), one quarter of the decay rate of Eq.~\eqref{eq:clean_normalized} (green), $2\mathcal I_{\rm typ}$ (red); dashed, Eq.~\eqref{eq:gamma_born}; dotted, the clean IPR floor; dash-dotted (left), the kernel bound $\gamma\le|k|$. Bottom: the same two diagnostics divided by $\gamma$.
}
\label{fig:xi_comp}
\end{figure}

Figure~\ref{fig:xi_comp} shows the correlator rate approaching $4\gamma$ as $N_t/\xiloc$ grows, reaching it at the strongest disorder, where $\gamma$ also comes closest to the kernel bound $|k|$ (dash-dotted line in Fig.~\ref{fig:xi_comp}). At weak disorder the fit window is capped by the ring and the minimum-image profile flattens, biasing the fitted slope low; the correlator is an unbiased measurement of $\gamma$ for $N_t\gg\xiloc$.

\subsection{Equal-time response}
\label{sec:eta}

On the disordered ring $\Pi_Q=\sum_{ij}e^{iQ(i-j)}\Pi_{ij}$ with $Q=2\pi n/N_t$ remains well defined, although $Q$ no longer labels the modes, and
\begin{align}
\Pi_Q(k,0)
&=k^2\sum_{m,n}
\sinh^2\vartheta_{mn,k}
|F^{mn}(Q)|^2,
\nonumber\\
F^{mn}(Q)
&=\sum_i e^{iQi}(U_g)_{im}(U_g)_{in}.
\label{eq:PiQ_modes}
\end{align}
Expanding $e^{iQ(i-j)}$, the constant term vanishes by the sum rule \eqref{eq:sumrule} and the linear one by symmetry, so
\begin{align}
\int_0^\Lambda dk\,\Pi_Q(k,0)
&=Q^2\eta+O(Q^4),
\nonumber\\
\eta
&=-\frac{1}{2}\int_0^\Lambda dk \sum_{i,j} d_{ij}^2\, \Pi_{ij}(k,0),
\label{eq:eta_disordered}
\end{align}
with $d_{ij}=\min(|i-j|,N_t-|i-j|)$ the minimum-image separation; in the uniform array this returns $A_\Lambda(0)$ of Eq.~\eqref{eq:A_explicit} up to the factor $N_t/2\pi$ (Appendix~\ref{app:eta_bloch}). Denoting by $\eta(\sigma)$ the average of Eq.~\eqref{eq:eta_disordered} over realizations at disorder strength $\sigma$, at small disorder
\begin{equation}
\frac{\eta(\sigma)}{\eta(0)}
=1-c\sigma^2+O(\sigma^4),
\label{eq:eta_perturbative}
\end{equation}
where the computed suppression is independent of $N_t$ to within a few percent (Fig.~\ref{fig:eta_N}) and of $V$ to within 1\% over $V=0.3$--$0.05$.

\begin{figure}
\centering
\includegraphics[width=\linewidth]{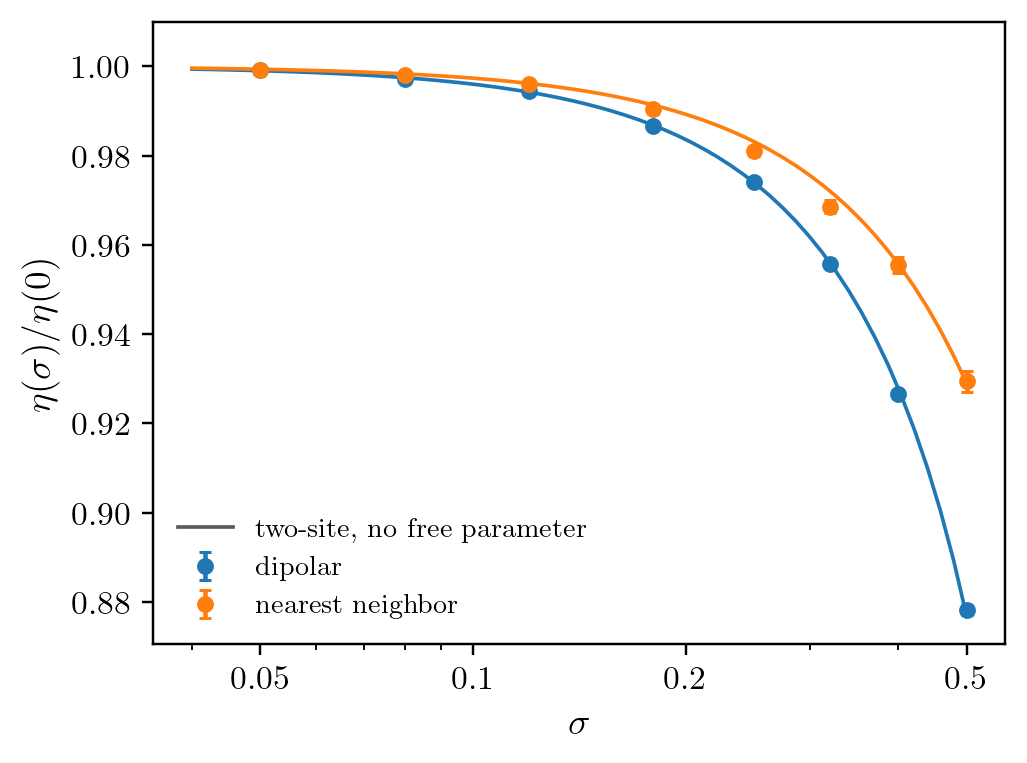}
\caption{Disorder dependence of the equal-time coefficient $\eta$, Eq.~\eqref{eq:eta_disordered}, at $N_t=1000$ and $V=0.3$. Points: full calculation, 100 realizations. Lines: the two-site average of Eq.~\eqref{eq:second_order}, with no free parameter. Cutoffs as in Appendix~\ref{app:num_cutoff}.}
\label{fig:eta_disorder}
\end{figure}

This weak, size- and coupling-independent suppression follows from perturbation theory in the inter-tube coupling. Taking the diagonal of $\hat G$, $g_i=n_iU_c$, as the unperturbed problem, each mode sits on a single tube, and the off-diagonal element $\hat G_{ij}$ mixes tubes $i$ and $j$ with amplitude $\hat G_{ij}/(g_i-g_j)$. The overlap \eqref{eq:pair_overlap} needs one such amplitude in each of its two modes, so the leading term of $\Pi_{ij}(k,0)$ is second order in $\hat G_{ij}$ and involves only the two tubes $i$ and $j$,
\begin{equation}
\Pi_{ij}(k,0)\simeq-2k^2\hat G_{ij}^2\,\frac{\sinh^2(\theta_{i,k}-\theta_{j,k})}{(g_i-g_j)^2},
\label{eq:second_order}
\end{equation}
with $g_i=n_iU_c$, $\hat G_{ij}=\sqrt{n_in_j}\,V_{ij}(k)$, and $\theta_{i,k}$ the squeeze angle of Eq.~\eqref{eq:def-uk-vk-hyperbolic} at $g=g_i$; the ratio stays finite as $g_i\to g_j$, where it agrees to the same order with the exact two-tube result. Through Eq.~\eqref{eq:eta_disordered}, $\eta$ is then a sum over pairs, weighted by $d_{ij}^2V_{ij}(k)^2$, of a function of $n_i$ and $n_j$ alone, and its disorder average gives Eq.~\eqref{eq:eta_perturbative} with $c$ independent of $V$ and of $N_t$ by construction. Evaluated with the distribution \eqref{eq:halfwidth}, at the cutoffs of Appendix~\ref{app:num_cutoff}, it gives $c=0.40$ (dipolar) and $0.27$ (n.n.), and with no free parameter it follows the computed $\eta(\sigma)/\eta(0)$ over the whole disorder range, Fig.~\ref{fig:eta_disorder}, including the $O(\sigma^4)$ correction that steepens the dipolar curve. The equal-time response is set by the interaction range because it is the momentum profile before anything has propagated; the localization length does not enter.

\subsection{Transverse momentum spreading}
\label{sec:no_transport}

\begin{figure}
\centering
\includegraphics[width=\linewidth]{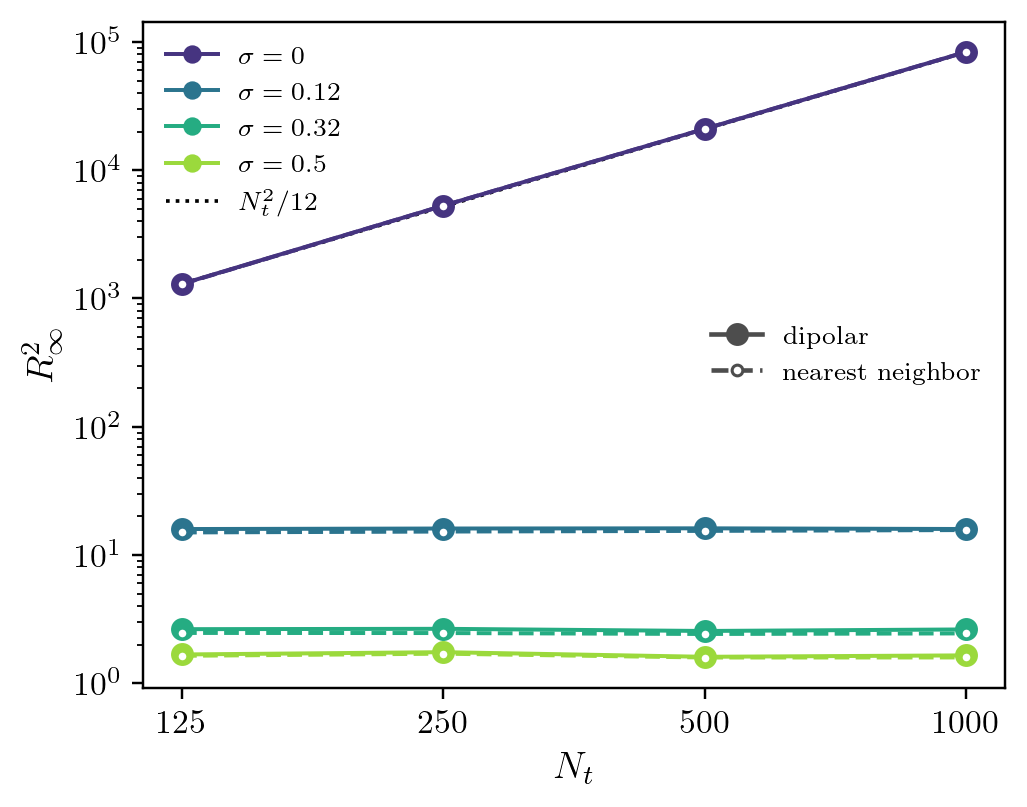}
\caption{Saturated transverse spread $R_\infty^2$, Eq.~\eqref{eq:R2inf}, against $N_t$ at $k=1$; color gives $\sigma$. Solid, filled: dipolar at $V=0.3$; dashed, open: n.n. at the matched bond $0.181$. Dotted: $N_t^2/12$.
}
\label{fig:momentum_spread}
\end{figure}

The saturated spread of the momentum profile is the long-time limit of the spatial variance \eqref{eq:R2_general}, with numerator and denominator averaged separately,
\begin{equation}
R_\infty^2(\sigma,N_t)
=\frac{
\displaystyle\sum_{j\neq i}d_{ij}^2
\left\langle|\Pi_{ij}|^2\right\rangle_t
}{
\displaystyle\sum_{j\neq i}
\left\langle|\Pi_{ij}|^2\right\rangle_t
},
\label{eq:R2inf}
\end{equation}
with the same long-time average as Sec.~\ref{sec:disorder-quench}. In the uniform array the dephased profile is uniform on the ring,
\begin{equation}
R_\infty^2\simeq\frac{N_t^2}{12},
\label{eq:R2_balanced}
\end{equation}
for either kernel; at any finite $\sigma$ it saturates at an $N_t$-independent value that decreases with $\gamma$, Fig.~\ref{fig:momentum_spread}, reached from the ballistic law of Sec.~\ref{sec:spreading} at $t\sim\xiloc/v_b$ (Fig.~\ref{fig:R2_time}). The spread is ballistic on the ring in the uniform array and arrested at $\xiloc$ when disordered. In both cases it is propagation in the pair channel of Sec.~\ref{sec:viscosity}; a dc momentum conductivity would need the intraband channel $E_m-E_n$, which opens at finite temperature with the same overlaps $\mathcal M^{mn}_{i,k}$, and we do not pursue it here.

\section{Instability of the uniform condensate}
\label{sec:instability}

\begin{figure}
\centering
\includegraphics[width=\linewidth]{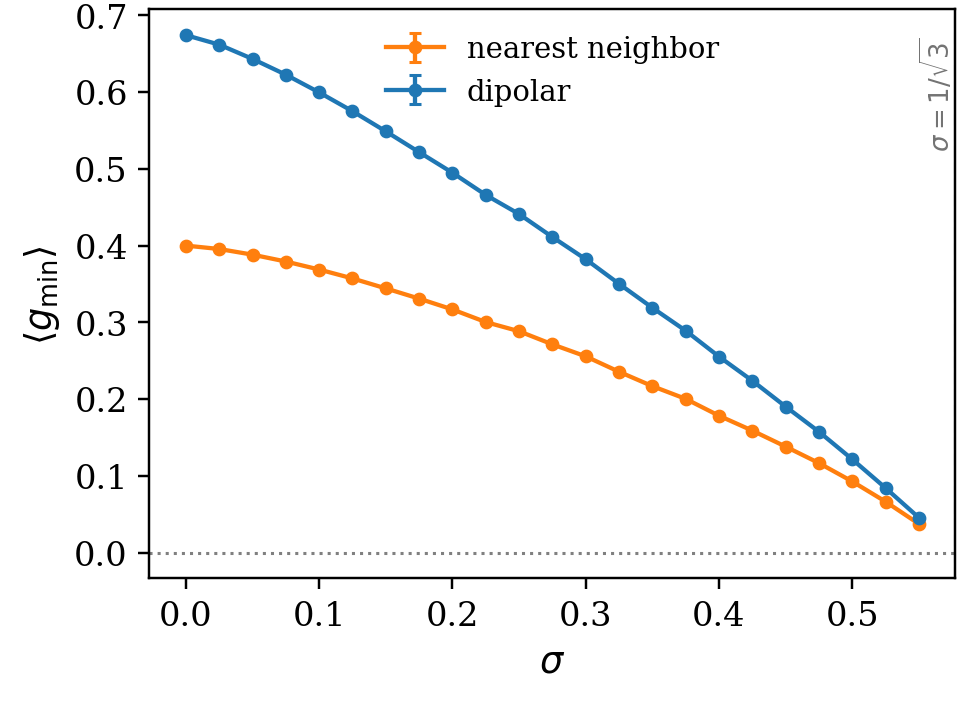}
\caption{Disorder-averaged minimum eigenvalue of $\hat G$ at $V=0.3$, $k=1$, $N_t=1000$, 100 realizations.}
\label{fig:gmin_softening}
\end{figure}

By Sec.~\ref{sec:stability} the stability of the uniform condensate is decided by the interaction matrix of the uniform array. Its minimum eigenvalue is at $q=\pi$ for both kernels,
\begin{equation}
g_\pi=n_0(U_c-2V)
\label{eq:gpi_nn}
\end{equation}
for n.n. coupling and, at $k\to0$,
\begin{equation}
g_\pi = n_0\left(U_c - \frac{\pi^2}{6}V\right)
\label{eq:gpi_dip}
\end{equation}
for dipolar coupling, from the alternating sum $2\sum_r(-1)^r/r^2=-\pi^2/6$. The critical couplings are
\begin{equation}
V_c^{\rm nn}=\frac{U_c}{2},
\qquad
V_c^{\rm dip}=\frac{6U_c}{\pi^2}\simeq0.608\,U_c ,
\label{eq:instability_thresholds}
\end{equation}
the dipolar one being the collapse threshold of the Gaussian sliding-liquid theory of Ref.~\cite{kollath}. The long-range kernel tolerates the larger coupling because tubes at even separation sit on the same sublattice of the $q=\pi$ pattern and stabilize it, which is why the alternating sum, $-1.64$, is smaller in magnitude than the $-2$ of a single bond. The dipolar $g_\pi(k)$ rises with $k$ (the threshold is $0.703\,U_c$ at $k=0.5$), so the first unstable mode is at $k\to0$, $q=\pi$. Density imbalance lowers the smallest eigenvalue, by an order of magnitude across the accessible range (Fig.~\ref{fig:gmin_softening}), without ever changing its sign, as Eq.~\eqref{eq:congruence} requires; the threshold is the same at every $\sigma$ to four digits.

\subsection{Unstable modes and the state beyond the threshold}

For $g_\pi<0$ the modes with
\begin{equation}
\varepsilon_k<-2g_\pi(k),\qquad\text{i.e.}\quad k^2<-4g_\pi ,
\label{eq:imaginary_E}
\end{equation}
have imaginary $E_{\pi,k}$ and grow at the rate $\Gamma_k=\sqrt{\varepsilon_k(-\varepsilon_k-2g_\pi)}$, which vanishes at $k\to0$ and is largest at $\varepsilon_k=-g_\pi$, $\Gamma_{\max}=|g_\pi|$ (for the dipolar kernel, with $g_\pi(k)$ rising with $k$, the maximum sits at somewhat smaller $k$). The unstable pattern is therefore two-dimensional in the $(x,i)$ plane: a longitudinal density wave of wavevector $k_*\simeq\sqrt{-2g_\pi}$ in each tube, antiphase between neighboring tubes. A purely transverse modulation, $k=0$ with alternate tubes fuller and emptier, is not available, since each $\hat N_i$ is conserved; the transverse $q=\pi$ pattern can only be realized by redistributing density along the tubes. The quadratic theory says nothing about the final state. The natural endpoint, with every $k<\sqrt{-4g_\pi}$ unstable, is coarsening to phase separation along each tube into dense and dilute segments, staggered between neighbors, as in the miscible--immiscible transition of a two-component condensate to which the problem reduces at $N_t=2$~\cite{ho1996,sabbatini2011,wilson2012} (threshold $V=U_c$ there, $U_c/2$ in the extended array because both neighbors soften the $q=\pi$ mode). Whether a periodic density wave at $k_*$ can instead be stabilized requires the terms dropped from Eq.~\eqref{eq:quadratic_hamiltonian}; Ref.~\cite{kollath} describes the state past the collapse as bunching within the tubes. The modulation energy of such a state is a Gross--Pitaevskii problem in the extended model, not a refinement of this one.

A modeling caveat: $U_c$ here is purely intra-tube and $V$ purely inter-tube, so $V/U_c$ is an anisotropy ratio. A physical dipolar gas also has a dipolar contribution to the intra-tube coupling~\cite{kollath}, which shifts $U_c$ and hence the threshold in terms of the bare dipolar strength; the ``dipolar stronger than contact'' regime is in any case not reached, since the array orders at $V_d\simeq0.30\,U_c$.

\section{Outlook}
\label{sec:discussion}
We have developed a novel approach to characterize
transverse momentum transport using both standard average correlator (termed ``amplitude'') and its variance (``intensity''). Both can be obtained in an experimental setup where momentum is imparted locally to one tube, with other tubes measured at later times, shot by shot, so that the same set of shots can be used to obtain both the amplitude and the intensity. Since each loading is a new realization of the density imbalance, the shot average performs the disorder average, and the two quantities measure the two sides of the paper: the amplitude obeys the sum rule and is suppressed by disorder, the intensity carries the ballistic spread $v_b^2t^2$ of the uniform array and its saturation at $\xiloc$ in the disordered one. On the theory side, the most immediate extension is to include heretofore neglected effects of finite temperature and interchain hopping. We expect a more conventional ``normal'' viscosity to emerge from incoherent hopping of excitations.

\begin{acknowledgments}
Claude (Anthropic; Opus 5.5 and Fable 5.1) was used throughout this work: to debug and extend the authors' numerical routines, to write most of the analysis and plotting code, to assist with derivations, including the closed form of the uniform-array correlator and the stationary-phase analysis of Sec.~\ref{sec:spreading}, to check formulas, and to draft and restructure parts of the text. Figures are plots of computed data produced by this code. The authors directed and verified all AI-assisted output; the numerical and analytical checks are described in Appendix~\ref{app:numerics}.
\end{acknowledgments}

\appendix

\section{Time and length scales}
\label{sec:timelengthscales}

Numbers are for $^{162}$Dy in the lattice of Ref.~\cite{tang2018}: $\lambda=741$~nm ($E_R=h^2/2M\lambda^2=h\times2.24$~kHz), lattice depth $s\equiv V_0/E_R=18$ unless stated, transverse confinement $\omega_\perp=2\sqrt{s}E_R/\hbar=2\pi\times19$~kHz, $a_s=100\,a_0$, $n_0=3\,\mu{\rm m}^{-1}$, $\mu=9.9\,\mu_B$, dipoles polarized normal to the plane of the array. Order-unity factors are tracked only where stated.

\begin{enumerate}

\item Healing length $\xi\equiv\hbar/\sqrt{Mn_0U_c}\simeq0.32\,\mu$m; sound velocity $v=\sqrt{n_0U_c/M}\simeq1.2$~mm/s; $\xi/v=\hbar/n_0U_c$.

\item Intra-tube interaction energy $g_U\equiv n_0U_c=2\hbar\omega_\perp a_sn_0=h\times600$~Hz (Lieb--Liniger $\gamma\simeq1.1$). The short time scale is $t_{\rm short}\equiv\hbar/g_U=0.26$~ms: the phonon at $k=1/\xi$ has $E_k=\sqrt{\varepsilon_k(\varepsilon_k+2g_U)}=1.12\,g_U$ with $\varepsilon_k=g_U/2$, so $t_{\rm short}$ is $1/2\pi$ of its period, and the crossover from linear to quadratic dispersion sits at $k\xi\simeq2$.

\item Tube spacing $a=\lambda/2=371$~nm$\,=1.15\,\xi$. This is not a coincidence of one experiment: for a single retro-reflected lattice $a/\xi=\pi\sqrt{2\sqrt{s}\,a_sn_0}$, independent of $\lambda$.

\item Nearest-tube bond $g_V\equiv Vn_0=2V_dn_0/a^2=h\times56$~Hz, so $g_V/g_U=0.093$; $t_V\equiv\hbar/g_V=2.9$~ms, $\tau_V\equiv t_V/t_{\rm short}=10.8$. At $k=1/\xi$ the bond is reduced by $(ka)K_1(ka)=0.54$, to $h\times30$~Hz. The ratio $V/U_c=4MV_d/(\sqrt{s}\,h^2a_s)$ is likewise independent of $\lambda$; it is set by $s$ and $a_s$. The transverse dispersion of the collective modes is generated by $g_V$ through $g_q(k)=g_U+2n_0\sum_rV_r(k)\cos qr$.

\item Transverse dispersion. As $k\to0$, $E_{q,k}=k\sqrt{g_q/M}$ with $g_0=g_U+(\pi^2/3)g_V$ and $g_\pi=g_U-(\pi^2/6)g_V=0.85\,g_U$, so $\Delta\omega(k)\equiv E_{0,k}-E_{\pi,k}=0.22\,kv$. At $k=1/\xi$, with the Bessel-reduced kernel and the full Bogoliubov form, $\Delta\omega=0.09\,g_U/\hbar=2\pi\times54$~Hz, giving $t_\Delta\equiv1/\Delta\omega=2.9$~ms$\,\simeq t_V$: the transverse hop time per tube is $\hbar/g_V(k)$.

\item Single-boson tunneling $J$ between tubes, the $t_\perp$ of Eq.~\eqref{eq:no_tunneling}, from the deep-lattice (Mathieu) asymptotics $J/E_R\simeq(4/\sqrt\pi)\,s^{3/4}e^{-2\sqrt s}$: $J/h=9.0$~Hz at $s=18$ (the value quoted in Ref.~\cite{tang2018}) and $0.020$~Hz at $s=60$.

\item Josephson frequency of the Bogoliubov mean-field state. With hopping, $\hat T$ acquires $\varepsilon_\perp(q)=2J(1-\cos q)$ and the $k=0$, $q=\pi$ mode has $E^2=\varepsilon_\perp(\varepsilon_\perp+2g_\pi)\simeq8Jg_\pi$, so $\omega_J=\sqrt{8Jg_\pi}/\hbar$, i.e. $\omega_J/2\pi=\sqrt{8\,(J/h)(g_\pi/h)}\simeq2.6\sqrt{Jg_U}/h$ (factor $16$ in place of $8$ for a square array). Values: $190$~Hz at $s=18$, $9$~Hz at $s=60$. The exponential suppression of $J$ enters only through its square root.

\item Infrared cutoff on $k$. Setting $\Delta\omega(\Lambda_{\rm IR})=\omega_J$ with the small-$k$ form of item~5 gives $\Lambda_{\rm IR}\xi=\hbar\omega_J/(0.22\,g_U)$: $1.4$ at $s=18$ and $0.07$ at $s=60$. (Comparing instead the hopping correction to $E_{q,k}$, which grows as $\varepsilon_\perp g_q/kv$ toward small $k$, with $\Delta\omega$ gives the same crossover within a factor $1.5$.) At $s=18$ the entire phonon regime, including $k\xi=1$, is transversely locked.

\item Self-consistency of $t_\perp=0$ requires $\Lambda_{\rm IR}\xi\ll1$. Demanding $\Lambda_{\rm IR}\xi=0.1$ gives $\omega_J/2\pi=13$~Hz, $J/h=0.04$~Hz, $J/E_R=1.9\times10^{-5}$, i.e. $s\simeq54$. Since $\Lambda_{\rm IR}\xi\propto\sqrt{J/g_U}$, lowering $g_U$ to reach the Bogoliubov regime raises the required depth ($\gamma\simeq0.3$ at this density needs $s\simeq58$).

\item Disorder scale. Density imbalance enters $\hat G$ as $\sigma g_U$ on the diagonal and $\sigma g_V(k)$ on the bonds. The fixed-$k$ localization results additionally require $4J\ll\sigma g_V(k)$: at $s=18$ this is $36$~Hz against $30\sigma$~Hz and fails for $\sigma\lesssim1$; at $s=60$ it holds comfortably.

\item Temperature. At $T>0$ the inter-tube phase locking of items 7--9 survives only below the temperature at which the Josephson coupling orders the array. In the random-phase approximation of Refs.~\cite{ho2004,cazalilla2006},
\begin{equation*}
\left(\frac{2\pi k_BT_c}{\hbar vn_0}\right)^{2-1/2K}=F(K)\,\frac{zJ}{\hbar vn_0},
\end{equation*}
with $F(K)=\pi^2\mathcal A_B(K)\sin(\pi/4K)\,B^2(1/8K,\,1-1/4K)$, $K=\pi\hbar n_0/Mv\simeq3.0$ the Luttinger parameter, $\mathcal A_B\simeq(K/\pi)^{1/2K}$, $B$ the beta function and $z$ the number of neighboring tubes; for $K\gg1$ it reduces to $k_BT_c\simeq2\pi\sqrt{zJ\hbar^2n_0^2/M}$. Above $T_c$ the array is incoherently coupled, and the remaining condition is that $k_BT$ stay below the collective-mode energies probed, $h\times600$~Hz $=30$~nK at $k\xi=1$. For a chain of tubes ($z=2$), $k_BT_c=h\times750$~Hz $=36$~nK at $s=18$, above that scale, so at this depth there is no window in which the zero-temperature results with $t_\perp=0$ apply. At $s=60$, $k_BT_c=h\times27$~Hz $=1.3$~nK, and the window is $1.3\,{\rm nK}\lesssim T\lesssim30$~nK. Since $T_c$ falls as $\sqrt J$, it drops below $30$~nK only for $s\gtrsim20$; at $s\simeq40$, $k_BT_c\simeq5$~nK, and the range is $5\,{\rm nK}\lesssim T\lesssim30$~nK.

\end{enumerate}

\section{Derivation of the momentum correlator}
\label{app:derivation}

We begin from Eq.~\eqref{eq:def-P-Mmn} and use the time evolution in Eq.~\eqref{eq:b_time_evolution}. At zero temperature the contractions needed for the four-operator expectation value include
\begin{align}
\langle
\beta_{m,-k}\beta_{n,-k}^\dagger
\beta_{m',-k'}\beta_{n',-k'}^\dagger
\rangle
&=\delta_{mn}\delta_{m'n'},
\label{ag1}
\\
\langle
\beta_{m,-k}\beta_{n,k}
\beta_{m',k'}^\dagger\beta_{n',-k'}^\dagger
\rangle
&=\delta_{mn'}\delta_{nm'}\delta_{kk'}.
\label{ag2}
\end{align}

The first contraction produces a time-independent factorized term and contributes to the disconnected part of the momentum correlator, removed by the connected definition in Eq.~\eqref{eq:def_pi}. The second gives the dynamical contribution

\begin{equation}
\begin{split}
\Pi_{ij}(t)={}&\int dk\,k^2\sum_{m,n}\mathcal M_{i,k}^{mn}\mathcal M_{j,k}^{nm}e^{i\mathcal E_{mn,k}t}\\
&\times\left(u_m^2v_n^2+v_m^2u_n^2-2u_mv_mu_nv_n\right),
\end{split}
\label{eq:Pi_dyn_app}
\end{equation}

where all coherence factors are at the same $k$, which we suppress, and the terms arise from pair creation and annihilation in the two transverse modes. Using $u_m=\cosh\theta_m$ and $v_m=\sinh\theta_m$,

\begin{align}
&u_m^2v_n^2+v_m^2u_n^2-2u_mv_mu_nv_n\nonumber\\
&\qquad=(\sinh\theta_m\cosh\theta_n-\cosh\theta_m\sinh\theta_n)^2\nonumber\\
&\qquad=\sinh^2(\theta_m-\theta_n),
\label{eq:squeeze_difference_identity}
\end{align}

which reduces Eq.~\eqref{eq:Pi_dyn_app} to Eq.~\eqref{e26a}.

\section{Properties of $\mathcal M$}
\label{app:M_properties}

The definition in Eq.~\eqref{eq:def_M} gives
\begin{equation}
\mathcal M_{i,k}^{mn}=\mathcal M_{i,k}^{nm},
\label{eq:M_symmetry}
\end{equation}
so the matrix is symmetric in the two transverse-mode indices. Orthogonality of $U_g$ further gives
\begin{equation}
\sum_m\mathcal M_{i,k}^{mm}=1,
\qquad
\sum_i\mathcal M_{i,k}^{mm}=1.
\label{eq:M_sum_rules}
\end{equation}
The first identity resolves the weight of a physical tube over transverse modes; the second normalizes the tube weight of each mode and is what makes Eq.~\eqref{eq:ipr} an inverse participation ratio.

\section{Uniform-array correlator asymptotics}
\label{app:clean_factorized}

We recast Eq.~\eqref{eq:clean_realspace} before taking asymptotics. With $\Sigma_q=(\varepsilon_k+g_q)/2E_{q,k}$, Eqs.~\eqref{eq:sc_definitions} and \eqref{eq:h_definition} give $c_q-s_q=1$ and hence $C_r-S_r=\delta_{r,0}$, so the bracket of Eq.~\eqref{eq:clean_realspace} becomes a difference of squares, $\Sigma_r^2-H_r^2=(\Sigma_r-H_r)(\Sigma_r+H_r)$, whose factors are the transforms of
\begin{equation}
P_q=\Sigma_q-h_q=\frac{\varepsilon_k}{2E_{q,k}},
\qquad
M_q=\Sigma_q+h_q=\frac{E_{q,k}}{2\varepsilon_k},
\label{eq:PM_definition}
\end{equation}
using $E_{q,k}^2=\varepsilon_k(\varepsilon_k+2g_q)$ in the second. Treating $r=0$ separately,
\begin{equation}
\Pi_r(k,0)=2k^2\left[P_rM_r-\tfrac14\delta_{r,0}\right],
\label{eq:PM_form}
\end{equation}
exactly, for either kernel. Since $P_qM_q=1/4$ pointwise, Parseval gives the sum rule of Eq.~\eqref{eq:sumrule}. The two factors carry reciprocal powers of $E_{q,k}$, and the asymptotics of Eqs.~\eqref{eq:xi_clean_NN} and \eqref{eq:xi_clean_dip} reduce to their behavior near the singularity nearest the real axis.

\subsection{Contour integral}

Both $P_r$ and $M_r$ are Fourier coefficients of the form of Eq.~\eqref{eq:Xr_definition} at $t=0$,
\begin{equation}
X_r=\int_{-\pi}^{\pi}\frac{dq}{2\pi}\,e^{-iqr}x_q ,
\label{eq:contour_start}
\end{equation}
with $x_q$ standing for $P_q$ or $M_q$. Both are $2\pi$-periodic, $g_q$ being a cosine series in the integer tube separation, and both are analytic on the real axis, since stability gives $g_q>0$ and hence $E_{q,k}>0$ there for $k\neq0$. Writing $q_0=\varsigma_0-i\kappa$ for the branch point lying nearest the real axis, with $\kappa>0$ its depth below it, $x_q$ is then analytic in the strip $|\operatorname{Im}q|<\kappa$. The value of $\kappa$ is kernel dependent and is determined below.

Take the rectangle with corners $\pm\pi$ and $\pm\pi-i\lambda$, for any $\lambda<\kappa$. It encloses no branch point, and its vertical sides cancel by periodicity, so Cauchy's theorem equates the top and bottom edges,
\begin{equation}
X_r=\int_{-\pi-i\lambda}^{\pi-i\lambda}\frac{dq}{2\pi}\,e^{-iqr}x_q .
\label{eq:contour_shifted}
\end{equation}
Substituting $q=q'-i\lambda$ with $q'$ real gives $e^{-iqr}=e^{-iq'r}e^{-\lambda r}$, and the constant factor comes out of the integral:
\begin{equation}
X_r=e^{-\lambda r}\int_{-\pi}^{\pi}\frac{dq'}{2\pi}\,e^{-iq'r}\,x_{q'-i\lambda}.
\label{eq:contour_extracted}
\end{equation}
The remaining integrand is bounded, so $|X_r|\lesssim e^{-\lambda r}$ for every $\lambda<\kappa$: the decay rate is the distance to the nearest branch point. At the instability of Sec.~\ref{sec:instability} the strip closes, $\kappa\to0$, and the exponential decay is lost. Since $x_q$ is real on the real axis, its singularities are mirror images in that axis and occur in conjugate pairs. For $r>0$ the factor $e^{-\lambda r}$ of Eq.~\eqref{eq:contour_extracted} decays only for $\lambda>0$, so the contour can be moved downward alone and the relevant branch point is the one below the real axis, with $\operatorname{Im}q_0=-\kappa$. The case $r<0$ follows from the evenness of $\Pi_r$.

The power of $r$ follows from the type of the branch point, which for both kernels below is of square-root type. Let $x_q\simeq A(q-q_0)^{\alpha}$ near $q_0$, with $\alpha>-1$, and deform the real-axis contour of Eq.~\eqref{eq:contour_start} into a keyhole about the cut descending from $q_0$. The arc at its tip contributes $O(\epsilon^{\alpha+1})$ and drops out, while the two sides of the cut do not cancel: parametrizing $q=q_0-is$ with $s>0$, their difference is proportional to $\sin(\pi\alpha)s^{\alpha}$, and $e^{-iqr}=e^{-i\varsigma_0r}e^{-\kappa r}e^{-sr}$. The remaining $r$ dependence is therefore
\begin{equation}
\int_0^\infty\!ds\,s^{\alpha}e^{-sr}=\Gamma(\alpha+1)\,r^{-\alpha-1},
\label{eq:cut_integral}
\end{equation}
the factor $e^{-sr}$ confining the integral to $s\lesssim1/r$, so that a region of width $1/r$ with integrand of size $r^{-\alpha}$ produces the power. Collecting the factors,
\begin{equation}
|X_r|\sim r^{-\alpha-1}e^{-\kappa r}
\label{eq:branch_asymptotic}
\end{equation}
up to a constant, with corrections down by powers of $1/r$; the accompanying phase $e^{-i\varsigma_0r}$ plays no role below. Applied to Eq.~\eqref{eq:PM_form}, the exponential rates of $P_r$ and $M_r$ add and their algebraic exponents add, so that $\Pi_r\propto r^{-(\alpha_P+\alpha_M)-2}e^{-2\kappa r}$.

\subsection{Short range interactions}

Here $g_q=n_0(U_c+2V\cos q)$ is entire, so the only branch points are those of $E_{q,k}$, at the zeros of $\varepsilon_k+2g_q$. These satisfy
\begin{equation}
\cos q=-z_k,
\qquad
z_k=\frac{k^2/2+2n_0U_c}{4n_0V},
\label{eq:nn_branch_condition}
\end{equation}
and stability requires $g_\pi>0$, that is $U_c>2V$, so $z_k\geq U_c/2V>1$ and no real solution exists. Setting $q=\pi+iy$ and using $\cos(\pi+iy)=-\cosh y$ gives $\cosh y=z_k$, so that
\begin{equation}
q_0=\pi\pm i\kappa_E,
\qquad
\kappa_E=\operatorname{arccosh}z_k,
\label{eq:nn_branch_points}
\end{equation}
of which the lower controls $r>0$, with $\varsigma_0=\pi$. The zero is simple, since $2g'_{q_0}=-4n_0V\sin q_0=-4in_0V\sinh\kappa_E\neq0$, so $E_{q,k}\propto(q-q_0)^{1/2}$ and Eq.~\eqref{eq:PM_definition} gives
\begin{equation}
\alpha_P=-\tfrac12,
\qquad
\alpha_M=+\tfrac12 .
\label{eq:nn_exponents}
\end{equation}
Hence $P_r\sim r^{-1/2}e^{-\kappa_Er}$ and $M_r\sim r^{-3/2}e^{-\kappa_Er}$, and
\begin{equation}
\Pi_r(k,0)\propto r^{-2}e^{-2\kappa_Er}
\label{eq:nn_appendix_asymptotic}
\end{equation}
as stated in Eq.~\eqref{eq:xi_clean_NN}. The same rate follows from a special-function representation: $s_q$, $c_q$, and $h_q$ are all of the form $(A+B\cos q)^{\pm1/2}$ up to constants, whose Fourier coefficients are Legendre functions of the second kind,
\begin{equation}
S_r,\ C_r,\ H_r
\propto Q_{r-1/2}(z_k),
\label{eq:legendre_coeffs}
\end{equation}
and writing $z_k=\cosh\eta$ the large-order behavior
\begin{equation}
Q_\nu(\cosh\eta)
\simeq
\sqrt{\frac{\pi}{2\nu\sinh\eta}}
 e^{-(\nu+1/2)\eta}
\label{eq:legendre_asymptotic}
\end{equation}
reproduces $r^{-1/2}e^{-\kappa_Er}$ per coefficient.

\subsection{Dipolar interactions}

No such representation is available, and the branch point now lies in $g_q$ itself. Using $K_1(z)\simeq\sqrt{\pi/2z}\,e^{-z}$ in Eq.~\eqref{eq:dipolar_Vr}, the interaction matrix element at large separation is
\begin{equation}
V_r(k)\simeq2V_d\sqrt{\frac{\pi|k|}{2}}\;r^{-3/2}e^{-|k|r}.
\label{eq:dipolar_Vr_asymptotic}
\end{equation}
In $\sum_rV_r(k)\cos(qr)$, written as $\tfrac12\sum_rV_r(e^{iqr}+e^{-iqr})$, it is the first half that can diverge in the lower half plane. With $\tilde u=e^{iq}e^{-|k|}$ the divergent part is
\begin{equation}
\sum_{r\geq1}r^{-3/2}\tilde u^{\,r}=\operatorname{Li}_{3/2}(\tilde u),
\label{eq:dipolar_polylog}
\end{equation}
which converges for $|\tilde u|<1$, that is for $\operatorname{Im}q>-|k|$, and is singular only at $\tilde u=1$. That point requires $\operatorname{Re}q=0$ in addition to $\operatorname{Im}q=-|k|$, so the branch points are
\begin{equation}
q_0=\pm i|k|,
\qquad
\varsigma_0=0,
\label{eq:dipolar_branch_points}
\end{equation}
above and below the zone center rather than the zone corner. The expansion of the polylogarithm about $\tilde u=1$ is
\begin{equation}
\operatorname{Li}_{3/2}(\tilde u)=\zeta(\tfrac32)-2\sqrt{\pi}\,(-\ln\tilde u)^{1/2}+O(\ln\tilde u),
\label{eq:polylog_expansion}
\end{equation}
and $-\ln\tilde u\simeq-i(q-q_0)$ near the branch point, so that
\begin{equation}
g_q=g_{q_0}+\mathcal{A}\,(q-q_0)^{1/2}+\ldots
\label{eq:g_dipolar_singular}
\end{equation}
with $g_{q_0}$ finite because $\zeta(3/2)$ converges. This finiteness is what separates the two kernels. In the stable regime $\varepsilon_k+2g_{q_0}\neq0$, so $E_{q_0,k}\equiv E_0\neq0$ and
\begin{equation}
\begin{split}
E_{q,k}={}&E_0\left[1+\frac{2\mathcal{A}(q-q_0)^{1/2}}{\varepsilon_k+2g_{q_0}}\right]^{1/2}\\
\simeq{}&E_0+\frac{E_0\mathcal{A}}{\varepsilon_k+2g_{q_0}}(q-q_0)^{1/2},
\end{split}
\label{eq:E_dipolar_singular}
\end{equation}
so $E_{q,k}$ and $1/E_{q,k}$ acquire singular parts of the same type and
\begin{equation}
\alpha_P=\alpha_M=+\tfrac12 .
\label{eq:dipolar_exponents}
\end{equation}
Both transforms therefore behave as $r^{-3/2}e^{-|k|r}$, and
\begin{equation}
\Pi_r(k,0)\propto r^{-3}e^{-2|k|r}
\label{eq:dip_appendix_asymptotic}
\end{equation}
as stated in Eq.~\eqref{eq:xi_clean_dip}. The contrast with Eq.~\eqref{eq:nn_appendix_asymptotic} is entirely in the exponents: $P_q$ and $M_q$ carry opposite powers of $E_{q,k}$, so their singular exponents cancel when $E_{q,k}$ vanishes at $q_0$ and add when it does not.

Two limits are worth recording. The branch points of $E_{q,k}$ exist for the dipolar kernel as well, at the complex $q$ solving $\varepsilon_k+2g_q=0$; for the couplings used here they lie further from the real axis than Eq.~\eqref{eq:dipolar_branch_points}, and they would take over on approach to the instability of Sec.~\ref{sec:instability}, where $g_q$ approaches zero and $\kappa_E$ of Eq.~\eqref{eq:nn_branch_points} descends toward the axis. And as $k\to0$ the branch points of Eq.~\eqref{eq:dipolar_branch_points} themselves reach the real axis at $q=0$, where $g_q$ acquires the $|q|$ non-analyticity of Eq.~\eqref{eq:g_dipolar_smallk}; the exponential rate vanishes with $k$, which is why no exponential survives the longitudinal integration and Eq.~\eqref{eq:clean_kint_dip} is a pure power law.

\section{Long-time asymptotics}
\label{app:J_asymptotics}

The $t^{-3}$ law follows from the $k\to0$ endpoint of Eq.~\eqref{eq:A_explicit} for either kernel, and we give that argument first. As $k\to0$ we have $\varepsilon_k\to0$, $g_q(k)\to g_q(0)$, and $E_{q,k}\to c_qk$ with $c_q=\sqrt{g_q(0)}$, so the phase is linear in $k$ and has no stationary point while the amplitude vanishes as $k^2$. With the retarded prescription $t\to t-i0^+$,
\begin{equation}
\int_0^\infty\!dk\,k^2e^{2ic_qkt}=\frac{-2i}{(2c_qt)^3},
\label{eq:k_endpoint}
\end{equation}
and inserting this into Eq.~\eqref{eq:A_explicit} gives
\begin{equation}
\operatorname{Im}A(t)
\underset{t\rightarrow\infty}{\sim}
-\frac{1}{64t^3}
\int_0^{2\pi}dq\,
\frac{(g'_q)^2}{g_q^{7/2}},
\label{eq:ImA_asymptotic}
\end{equation}
which is Eq.~\eqref{eq:A_powerlaw}. The derivation assumes nothing about the kernel beyond $g_q(0)>0$ and the convergence of the $q$ integral, and both hold in either case. For nearest-neighbor coupling $g_q=n_0(U_c+2V\cos q)$ is smooth and bounded away from zero throughout the stable regime. For the dipolar kernel $g_q(0)$ is the piecewise-quadratic form of Eq.~\eqref{eq:g_dipolar_smallk}, likewise positive there, while $g'_q$ acquires a jump at $q=0$ from the $|q|$ term; it remains bounded, so the kink changes the coefficient and not the power. The integral diverges as $g_\pi\to0$, which is the threshold singularity discussed in Sec.~\ref{sec:viscosity}.

For nearest-neighbor coupling the reduction of Eq.~\eqref{eq:A_master} makes the same limit explicit. The phase $x\sqrt{1+x^2}$ of Eq.~\eqref{eq:A_master} likewise has no stationary point, and the same endpoint expansion gives $J(s,\infty)=-2i/s^3+O(s^{-5})$, which returns Eq.~\eqref{eq:ImA_asymptotic} on substitution. This is the argument above in the rescaled variable rather than a second derivation, and the independent checks are numerical. Direct evaluation of Eq.~\eqref{eq:A_master} approaches the coefficient $64$ from below as the subleading $t^{-5}$ correction dies off, reaching $63.99$ at $t=80$, and quadrature of Eq.~\eqref{eq:A_explicit} for the dipolar kernel reproduces the same power. The local exponent in Fig.~\ref{fig:clean_t3} is shown only to $t=50$: the coefficient is evaluated pointwise as $A(t)t^3$ and stays accurate well beyond that, whereas the numerical derivative of $\ln A$ amplifies roundoff once $|{\rm Im}\,A|$ falls to $10^{-8}$.

\section{Small-$Q$ interaction integrals}
\label{app:dipolar_integral}

For nearest-neighbor interactions we define
\begin{equation}
\alpha=n_0U_c,
\qquad
\beta=2n_0V,
\qquad
k_{\rm el}=\sqrt{\frac{2\beta}{\alpha+\beta}},
\label{eq:elliptic_parameters}
\end{equation}
in terms of which the transverse integral at $t=0$ is
\begin{equation}
\int_0^{2\pi}dq\,
\frac{(g'_q)^2}{\sqrt{g_q}}
=\frac{8}{3}\sqrt{\alpha+\beta}
\left[
\alpha\mathbb E(k_{\rm el})
-(\alpha-\beta)\mathbb K(k_{\rm el})
\right],
\label{eq:q_integral_NN}
\end{equation}
where $\mathbb K$ and $\mathbb E$ are complete elliptic integrals with parameter $m=k_{\rm el}^2$. The logarithmic divergence of $\mathbb K$ as $V\rightarrow U_c/2$ is cancelled by the factor $\alpha-\beta$, so the transverse integral stays finite at threshold and the singular behavior of $\eta$ there comes from the longitudinal integral instead.

For dipolar interactions $g_q(k)$ depends on both momenta. In the $k\rightarrow0$ limit,
\begin{equation}
\sum_{r=1}^\infty\frac{\cos(qr)}{r^2}
=\frac{\pi^2}{6}-\frac{\pi|q|}{2}+\frac{q^2}{4},
\qquad
q\in[-\pi,\pi],
\label{eq:dipolar_series}
\end{equation}
so that
\begin{equation}
g_q^{\rm dip}(0)
=n_0\left[
U_c+2V\left(
\frac{\pi^2}{6}-\frac{\pi|q|}{2}+\frac{q^2}{4}
\right)
\right],
\label{eq:g_dipolar_smallk}
\end{equation}
which is piecewise quadratic in $q$ and, evaluated at $q=\pi$, gives Eq.~\eqref{eq:gpi_dip} and the dipolar threshold of Eq.~\eqref{eq:instability_thresholds}. At finite $k$ the dipolar $g_q(k)$ has no closed form, and we evaluate the $q$ integral in Eq.~\eqref{eq:A_explicit} numerically.

\section{Momentum correlator in the tube basis}
\label{app:original_basis}

The time-evolved operators in the original tube basis are
\begin{equation}
a_{i,k}(t)
=\sum_m\left[
 u_{im,k}\beta_{m,k}e^{iE_{m,k}t}
+v_{im,k}\beta_{m,-k}^\dagger e^{-iE_{m,k}t}
\right],
\label{eq:tube_basis_time_evolution}
\end{equation}
with
\begin{equation}
u_{im,k}=(U_g)_{im}u_{m,k},
\qquad
v_{im,k}=(U_g)_{im}v_{m,k}.
\label{eq:tube_basis_coherence}
\end{equation}
Using the contractions of Appendix~\ref{app:derivation},
\begin{equation}
\Pi_{ij}(t)
=\int dk\,k^2\sum_{m,n}
 e^{i\mathcal E_{mn,k}t}\mathcal Q_{ij}^{mn},
\label{eq:orig_basis}
\end{equation}
where
\begin{align}
\mathcal Q_{ij}^{mn}
={}&u_{im}u_{jm}v_{in}v_{jn}
+v_{im}v_{jm}u_{in}u_{jn}
\nonumber\\
&-u_{im}v_{jm}u_{jn}v_{in}
-u_{jm}v_{im}u_{in}v_{jn}.
\label{eq:orig_basis_Q}
\end{align}
Substituting Eq.~\eqref{eq:tube_basis_coherence} reproduces the factorized overlap of Eq.~\eqref{eq:pair_overlap} and therefore Eq.~\eqref{e26a}.

\section{Small-$Q$ coefficient in the Bloch basis}
\label{app:eta_bloch}

Equation~\eqref{eq:eta_disordered} defines $\eta$ from the eigenbasis of $\hat G$, which is the form we evaluate with disorder. Here we check that it returns $A_\Lambda(0)$ of Eq.~\eqref{eq:A_explicit} when the array is uniform.

For $n_i=n_0$ the eigenvectors are Bloch waves, $(U_g)_{im}=e^{iq_mi}/\sqrt{N_t}$, so the overlap of Eq.~\eqref{eq:PiQ_modes} is
\begin{equation}
F^{mn}(Q)=\frac{1}{N_t}\sum_ie^{iQi}e^{i(q_n-q_m)i}=\delta_{Q,\,q_m-q_n},
\label{eq:F_bloch}
\end{equation}
and the double mode sum collapses to a single one,
\begin{equation}
\Pi_Q(k,0)=k^2\sum_q\sinh^2\!\left(\theta_{q+Q,k}-\theta_{q,k}\right),
\label{eq:PiQ_bloch}
\end{equation}
which is Eq.~\eqref{eq:PiQ_clean} at $t=0$. Expanding at small $Q$ gives $\sinh^2(\theta_{q+Q}-\theta_q)\simeq Q^2(\theta'_q)^2$. To obtain $\theta'_q$, differentiate $\cosh2\theta_{q,k}=(\varepsilon_k+g_q)/E_{q,k}$ from Eq.~\eqref{eq:def-uk-vk-hyperbolic}. Since $E_{q,k}^2=\varepsilon_k(\varepsilon_k+2g_q)$ gives $E'_{q,k}=\varepsilon_kg'_q/E_{q,k}$,
\begin{equation}
\frac{d}{dq}\cosh2\theta_{q,k}
=\frac{g'_q}{E_{q,k}}-\frac{(\varepsilon_k+g_q)\varepsilon_kg'_q}{E_{q,k}^3}
=\frac{\varepsilon_kg_qg'_q}{E_{q,k}^3},
\label{eq:dcosh}
\end{equation}
while $\sinh2\theta_{q,k}=\sqrt{\cosh^22\theta_{q,k}-1}=g_q/E_{q,k}$. Equating this to $2\theta'_q\sinh2\theta_{q,k}$ gives
\begin{equation}
\theta'_q=\frac{\varepsilon_kg'_q}{2E_{q,k}^2}=\frac{g'_q}{2(\varepsilon_k+2g_q)}.
\label{eq:theta_prime}
\end{equation}
Substituting into Eq.~\eqref{eq:PiQ_bloch} and taking $\sum_q\to N_t\int dq/2\pi$,
\begin{equation}
\eta = \frac{N_t}{8\pi}\int_0^\Lambda dk\, k^2 \int_0^{2\pi} dq\, \frac{(g_q')^2}{(\varepsilon_k + 2g_q)^2} = \frac{N_t}{2\pi}\,\mathrm{Re}\,A_\Lambda(0),
\label{eq:eta_bloch_final}
\end{equation}
the integrand being that of Eq.~\eqref{eq:A_explicit} at $t=0$. The factor $N_t/2\pi$ is the density of states converting the mode sum to the continuum $q$ integral, and is the normalization absorbed after Eq.~\eqref{eq:PiQ_clean}; it is independent of $\sigma$ and of the kernel, so it cancels in the ratio $\eta(\sigma)/\eta(0)$ that every result in Sec.~\ref{sec:eta} uses.

\section{Numerical methods}
\label{app:numerics}

Unless stated otherwise, the finite-array calculations use periodic boundary conditions, $N_t=1000$ tubes, $U_c=n_0=1$, and $100$ disorder realizations. All localization results are obtained at fixed longitudinal momentum, that is for the integrand of Eq.~\eqref{e26a} rather than the $k$-integrated correlator. We use the minimum-image distance
\begin{equation}
d_{ij}=\min(|i-j|,N-|i-j|),
\label{eq:minimum_image}
\end{equation}
which preserves translation invariance in the uniform finite ring. Each analytical reduction was checked against direct evaluation, and the implementation against its known limits: the uniform-array correlator, the sum rule, the congruence of Eq.~\eqref{eq:congruence}, the standing-wave IPR and the three-term recursion for nearest neighbors are reproduced, the threshold located by bisection is constant across $\sigma$ as Sec.~\ref{sec:stability} requires, and no realization was dynamically unstable.

\subsection{Momentum correlator}
\label{app:num_pi}

In the uniform array the transforms of Eq.~\eqref{eq:Xr_definition} are evaluated by FFT, so that Eq.~\eqref{eq:clean_realspace} returns the whole profile $\Pi_r$ at a cost of three transforms per $k$, against $O(N_t^4)$ for the direct double mode sum of Eq.~\eqref{e26a} over all tube pairs. This is how the uniform-array curves of Fig.~\ref{fig:clean_quench} are obtained. With disorder the same factorization is used in matrix rather than transform form. For each realization we construct $\hat G(k)$ and diagonalize it once. Defining
\begin{equation}
s_m=\sinh^2\theta_m,
\;
c_m=\cosh^2\theta_m,
\;
h_m=\sinh\theta_m\cosh\theta_m,
\label{eq:numerical_sch}
\end{equation}
the fixed-$k$ correlator follows from Eq.~\eqref{eq:squeeze_factorize} as
\begin{equation}
\Pi_{ij}(k,t)
=2k^2\left[A_{ij}(t)B_{ij}(t)-C_{ij}^2(t)\right],
\label{eq:num_pi}
\end{equation}
where
\begin{align}
A_{ij}(t)
&=\sum_m(U_g)_{im}s_me^{iE_{m,k}t}(U_g)_{jm},
\\
B_{ij}(t)
&=\sum_m(U_g)_{im}c_me^{iE_{m,k}t}(U_g)_{jm},
\\
C_{ij}(t)
&=\sum_m(U_g)_{im}h_me^{iE_{m,k}t}(U_g)_{jm}.
\label{eq:ABC_numerics}
\end{align}
The factorized form replaces the explicit double mode sum by three matrix products. For the infinite-time squared correlator we evaluate Eq.~\eqref{eq:timeavg} without a finite time grid: the identity $\sinh^2(\theta_m-\theta_n)=s_mc_n+c_ms_n-2h_mh_n$ expands into six terms, each a product of a function of $m$ with a function of $n$, so the double mode sum collapses into matrix products built from $|U_g|^2$ and the result is the $T\to\infty$ limit exactly.

\subsection{Averaging and fit windows}

We average spatial profiles geometrically over disorder realizations and over tube pairs at fixed separation, since the arithmetic mean is dominated by rare large-amplitude samples and does not converge as the window is pushed outward. We fit over
\begin{equation}
3\leq d/\xiloc\leq8,
\label{eq:fit_window}
\end{equation}
with $\xiloc=1/\gamma$. A window fixed in absolute distance cannot serve, since $\xiloc$ varies by nearly two orders of magnitude across the disorder grid: at weak disorder a window of a few tens of sites lies within a single localization length and is not asymptotic, while at strong disorder the squared correlator, decaying at $4\gamma$, has already reached the roundoff floor there. The window is clipped by the largest available separation and by the precision floor, which makes the weakest-disorder points the most sensitive to finite size. That bias is set by $N_t/\xiloc$ and not by the kernel. For nearest neighbors at $V=0.3$, where $\xiloc$ is about three times longer than at the matched coupling of Sec.~\ref{sec:localization}, the ratio at $\sigma=0.25$ is $0.78$ against $0.90$, and at the two weakest disorders the window cannot reach $3\xiloc$ at all; at $\sigma=0.50$, where $N_t/\xiloc$ is large at both couplings, the two agree within a few percent.

\subsection{Lyapunov exponent}
\label{app:num_tm}

At fixed $k$ we represent the eigenvalue problem for $\hat G(k)$ by a banded transfer matrix. With the off-diagonal coupling truncated at range $b$, the state vector is
\begin{equation}
x_i=(\psi_{i+b-1},\ldots,\psi_{i-b})^T,
\label{eq:transfer_state}
\end{equation}
so the transfer matrix has dimension $2b$ and Lyapunov exponents in positive and negative pairs. The localization rate is the smallest positive exponent, which controls the slowest asymptotic decay. We propagate an orthonormal frame with periodic QR reorthogonalization and accumulate the logarithms of the diagonal entries of $R$. Production calculations use $b=5$, chains of $1.5\times10^5$ sites, independent of the ring size $N_t$ used elsewhere, and four independent chains, whose spread gives the error bars in Fig.~\ref{fig:xi_comp}. At sufficiently small $k$ the dipolar coupling becomes effectively long ranged and a finite-band transfer matrix is no longer a controlled approximation.

\begin{figure*}
\centering
\includegraphics[width=1.4\columnwidth]{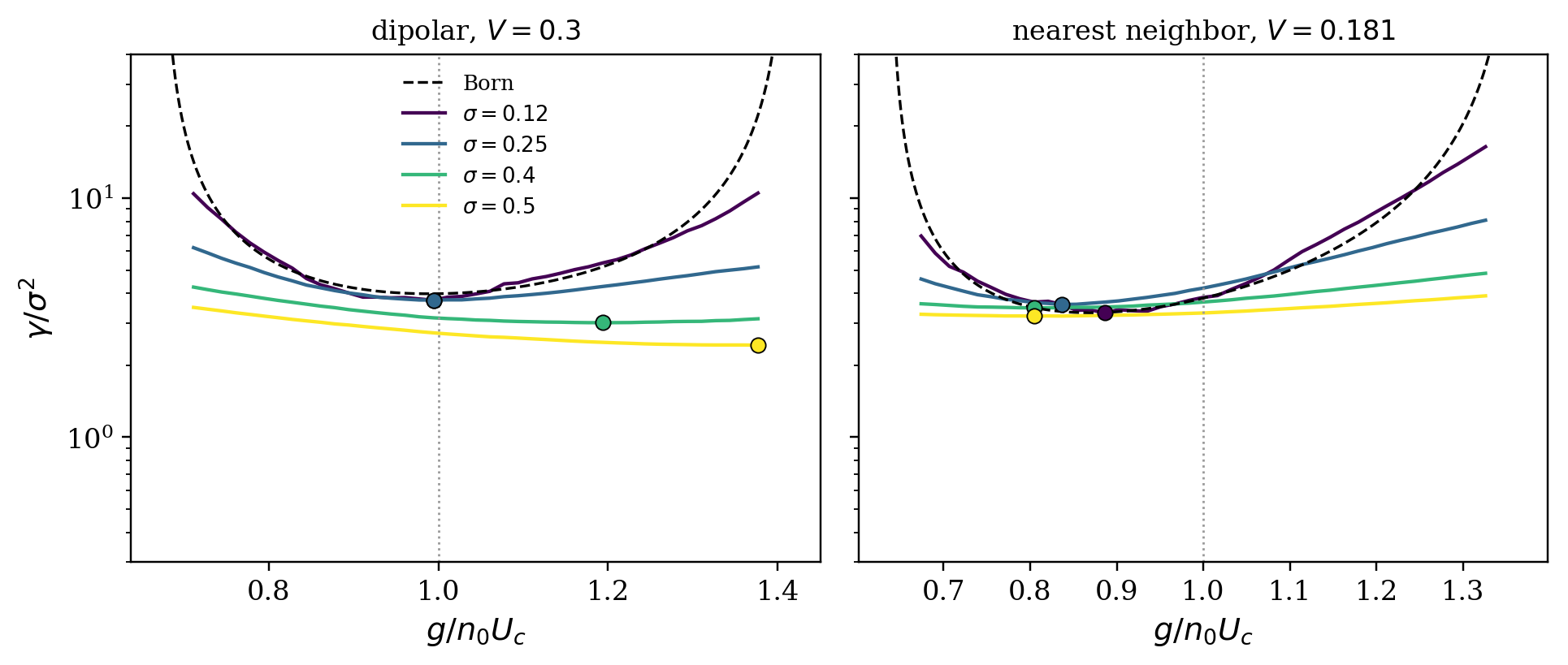}
\caption{Transfer-matrix Lyapunov exponent across the band at $k=1$, scaled by $\sigma^2$, for the dipolar kernel at $V=0.3$ (left) and for nearest neighbors at $V=0.181$, the matched first-neighbor bond of Sec.~\ref{sec:localization} (right). Dots mark the minimum at each $\sigma$, which is the rate the correlator measures, and the dotted line marks the band center $g=n_0U_c$, where the rate of Fig.~\ref{fig:V_dependence} is taken. The dashed line is the random-mass Born result of Eq.~\eqref{eq:gamma_random_mass}, with no free parameter, and the weakest disorder follows it across the band. The Born rate is the ratio of the random potential, which grows with $g$, to the squared group velocity. For nearest neighbors the velocity is symmetric about the band center, so the growing potential shifts the minimum below it. For the dipolar kernel at $k=1$ the band is asymmetric, and the asymmetry offsets this shift, placing the minimum at the band center throughout the Born regime; it leaves the center only once that regime ends. Each curve is one chain of $1.2\times10^5$ sites, with the same disorder sequence at every $g$.}
\label{fig:gamma_band}
\end{figure*}

The Lyapunov exponent depends on spectral position as well as on disorder strength, Fig.~\ref{fig:gamma_band}, and the dependence follows from the congruence of Eq.~\eqref{eq:congruence}. Writing $\hat G\psi=g\psi$ as $M\phi=gD^{-1}\phi$ with $\phi=D^{1/2}\psi$ leaves the disorder-free band of $M$ intact and turns the density fluctuations into a random potential $-(g/n_0)u_i$ to first order, whose strength grows with $g$; since $D^{1/2}$ is bounded, $\psi$ and $\phi$ share their Lyapunov exponent. The Born approximation for this potential gives
\begin{equation}
\gamma(g_q)\simeq\frac{\sigma^2g_q^2}{2\left(\partial_qg_q\right)^2},
\label{eq:gamma_random_mass}
\end{equation}
which reduces to Eq.~\eqref{eq:gamma_born} at $g=n_0U_c$ and follows the transfer matrix across the band at weak disorder. The potential grows with $g$ while the group velocity vanishes at both band edges, and the minimum of Eq.~\eqref{eq:gamma_random_mass} sits where the two balance: at $g^*=n_0(U_c-4V^2/U_c)$ for nearest neighbors, below the band center, and at $g\simeq n_0U_c$ for the dipolar band at $k=1$.

Two rates therefore enter the paper, and each comparison uses the one it tests. Figure~\ref{fig:V_dependence} tests Eq.~\eqref{eq:gamma_born}, a statement at fixed energy, and uses the rate at the band center $g=n_0U_c$ for both kernels; for nearest neighbors this is $q=\pi/2$, where the Kappus--Wegner anomaly~\cite{kappus1981} lowers the rate below Eq.~\eqref{eq:gamma_born} by $9\%$ as $\sigma\to0$. At finite $\sigma$ it competes with the nonlinearity of the random potential, whose exact form $g(1/n_i-1/n_0)$ has a variance exceeding $g^2\sigma^2/n_0^2$ by a term of relative order $\sigma^2$, so the band-center rate departs from Eq.~\eqref{eq:gamma_born} by a few percent in either direction until the Born regime itself ends. Figures~\ref{fig:clean_quench}(c) and \ref{fig:xi_comp} compare with the long-time correlator, whose tail is a sum over all modes and therefore decays at the smallest rate in the band. There we scan the band at each $\sigma$ with chains of $2\times10^4$ sites, locate the minimum positive exponent, and evaluate it with the production chains. The two rates coincide for the dipolar kernel at weak disorder and differ for nearest neighbors, where the band-center rate is the larger; this is why the nearest-neighbor minimum lies below the band-center Born line of Eq.~\eqref{eq:gamma_born}.

\subsection{Eigenvector envelopes}
\label{app:num_env}

\begin{figure}
\centering
\includegraphics[width=\linewidth]{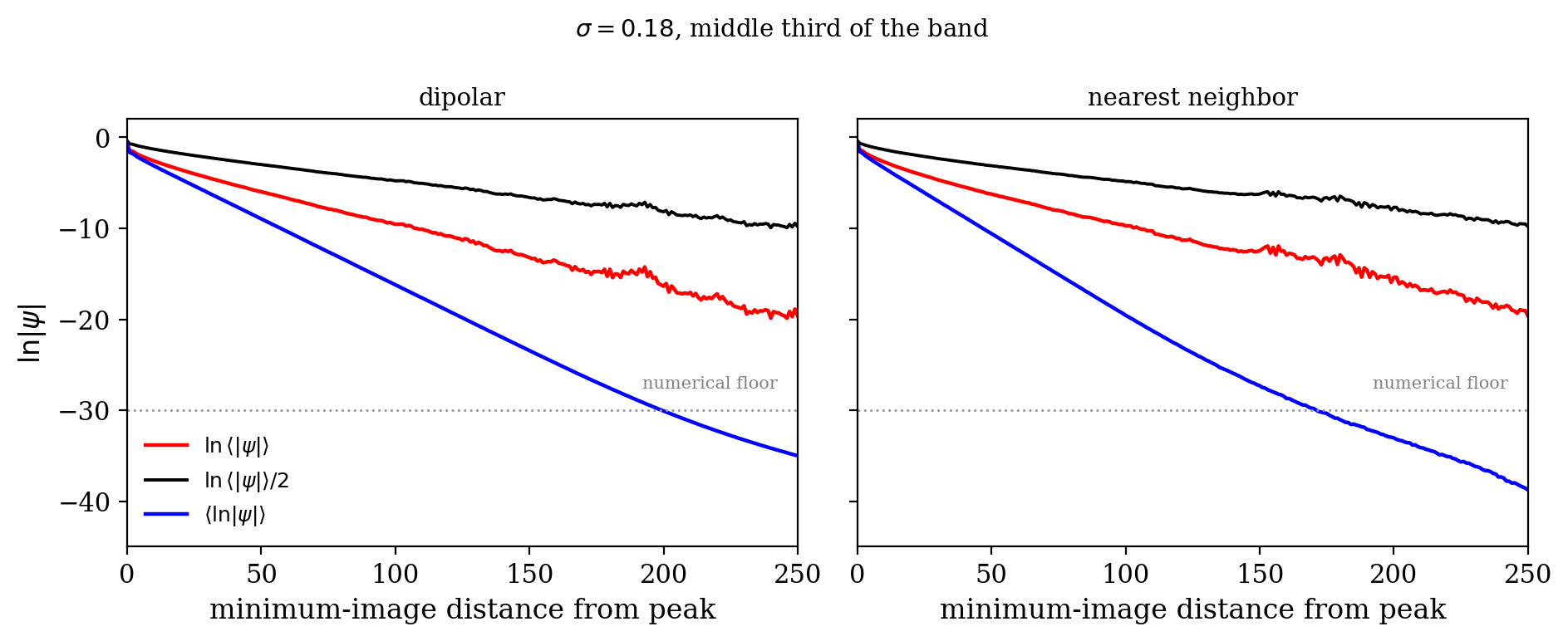}
\caption{Typical (blue) and annealed (red) eigenvector envelopes against the distance from the mode maximum, $\sigma=0.18$, $k=1$, $N_t=1000$, 100 realizations, middle third of the band; black, half the annealed envelope. Dipolar at $V=0.3$ (left) and n.n. at the matched bond $0.181$ (right). Dotted: numerical (roundoff) floor.
}
\label{fig:metrics}
\end{figure}

Figure~\ref{fig:metrics} compares two averages of the eigenvector amplitude at distance $d$ from the mode maximum, the typical envelope $\langle\ln|\psi|\rangle$ and the annealed envelope $\ln\langle|\psi|\rangle$. The amplitude at fixed $d$ is log-normally distributed, so the annealed average is dominated by rare large amplitudes and lies above the typical one by an amount that grows with $d$; only the typical envelope decays at $\gamma$. The IPR, a fourth moment of $|\psi|$, weights the same upper tail, which is one source of the drift of $2\mathcal I_{\rm typ}/\gamma$ in Fig.~\ref{fig:xi_comp}.

\subsection{Evaluation of $A(t)$ and $\eta$}
\label{app:num_cutoff}

The function $A(t)$ enters only Fig.~\ref{fig:clean_t3} and is a uniform-array quantity. For nearest neighbors we use the closed form of Eq.~\eqref{eq:A_master}, which integrates to $\Lambda\to\infty$; for the dipolar kernel we evaluate Eq.~\eqref{eq:A_explicit} by Gauss--Legendre quadrature in $q$ and $k$. The $t^{-3}$ law places a specific demand on the cutoff. It is an asymptotic statement about the $k\to0$ endpoint, and a finite cutoff leaves an endpoint term oscillating as $e^{2iE(\Lambda)t}$ that decays more slowly than $t^{-3}$, so the integrand must be negligible at $\Lambda$ rather than merely truncated there. The dipolar integrand falls as $e^{-2k}$ and is negligible at $\Lambda=10$, beyond which $\alpha(t)$ no longer moves.

The coefficient $\eta$ is evaluated with disorder, realization by realization, in the eigenbasis of $\hat G(k)$: we compute $\Pi_Q(k,0)$ from Eq.~\eqref{eq:PiQ_modes} at the smallest ring wavevector $Q=2\pi/N_t$, divide by $Q^2$, and integrate over $k$ by Gauss--Legendre quadrature, normalized so that the uniform array returns $A_\Lambda(0)$ of Eq.~\eqref{eq:A_explicit}. For nearest neighbors $\hat G$ is independent of $k$ and one diagonalization per realization serves every quadrature node; for the dipolar kernel $\hat G(k)$ is diagonalized at each node.

The cutoff for $\eta$ is set by the large-$k$ behavior of its integrand, which is fixed by the kernel rather than by the disorder. For $\varepsilon_k\gg g_m$ the squeeze angles are small, $\theta_{m,k}\simeq g_m/k^2$, so $\sinh^2\vartheta_{mn,k}\simeq(g_m-g_n)^2/k^4$, and the mode sum of Eq.~\eqref{eq:PiQ_modes} collapses to a commutator with the transverse phase,
\begin{align}
\sum_{m,n}(g_m-g_n)^2\left|F^{mn}(Q)\right|^2
&=\sum_{i,j}\hat G_{ij}^2\left|e^{iQi}-e^{iQj}\right|^2
\nonumber\\
&\simeq Q^2\sum_{i\neq j}d_{ij}^2\,V_{ij}(k)^2\,n_in_j,
\label{eq:tail_general}
\end{align}
for any density profile. The large-$k$ integrand of $\eta$ therefore falls as $k^{-2}$ with a $k$ dependence set by the kernel alone, and since distinct tubes are independent, $\langle n_in_j\rangle=n_0^2$ for $i\neq j$ and its disorder average equals the uniform-array value, which in the form of Eq.~\eqref{eq:A_explicit} is $4(g'_q)^2/k^2$. For nearest neighbors $g_q=n_0(U_c+2V\cos q)$ carries no $k$, and
\begin{equation}
\int_\Lambda^\infty dk\,\frac{4(g'_q)^2}{k^2}=\frac{4(g'_q)^2}{\Lambda},
\label{eq:tail_NN}
\end{equation}
an algebraic tail. For the dipolar kernel $g'_q=-2n_0Vk\sum_rK_1(kr)\sin qr$ from Eq.~\eqref{eq:dipolar_fourier}, and $K_1(z)\to\sqrt{\pi/2z}\,e^{-z}$ leaves the $r=1$ term dominant, so $g'_q\sim e^{-k}$ and
\begin{equation}
\int_\Lambda^\infty dk\,\frac{4(g'_q)^2}{k^2}\sim\frac{e^{-2\Lambda}}{2\Lambda^2}.
\label{eq:tail_dip}
\end{equation}
The dipolar truncation is therefore negligible at modest $\Lambda$, and $\Lambda=6$ suffices, while the nearest-neighbor one is reduced only in proportion to $1/\Lambda$. For nearest neighbors the exact $\Lambda\to\infty$ value is known in the uniform array, since $\int_0^\infty dk\,k^2(\varepsilon_k+2g_q)^{-2}=\pi/2\sqrt{g_q}$ gives $\eta=\tfrac{\pi}{8}\int dq\,(g'_q)^2/\sqrt{g_q}$, the elliptic integral of Eq.~\eqref{eq:q_integral_NN}, and the computed $\eta$ at $\sigma=0$ approaches it with the $1/\Lambda$ deficit that Eq.~\eqref{eq:tail_NN} predicts. That value cannot set the cutoff, being a uniform-array result with no counterpart at finite $\sigma$. The cutoff is instead set on the quantity we report: $\Lambda$ is raised until the suppression $1-\eta(\sigma)/\eta(0)$ stops moving, which happens well before $\eta$ itself has converged, since by Eq.~\eqref{eq:tail_general} the truncated tail is common to numerator and denominator on average, to leading order in $g/\varepsilon_k$. We take $\Lambda=800$ for nearest neighbors and split the quadrature at $k=10$ to keep the small-$k$ region resolved. The same fact sets how the two kernels weight the disorder: the nearest-neighbor $k^{-2}$ tail places much of $\eta$ at large $k$, where its average is independent of $\sigma$, so the suppression is diluted and $c$ is smaller, while the dipolar integrand never reaches that regime.

\begin{figure}
\centering
\includegraphics[width=\linewidth]{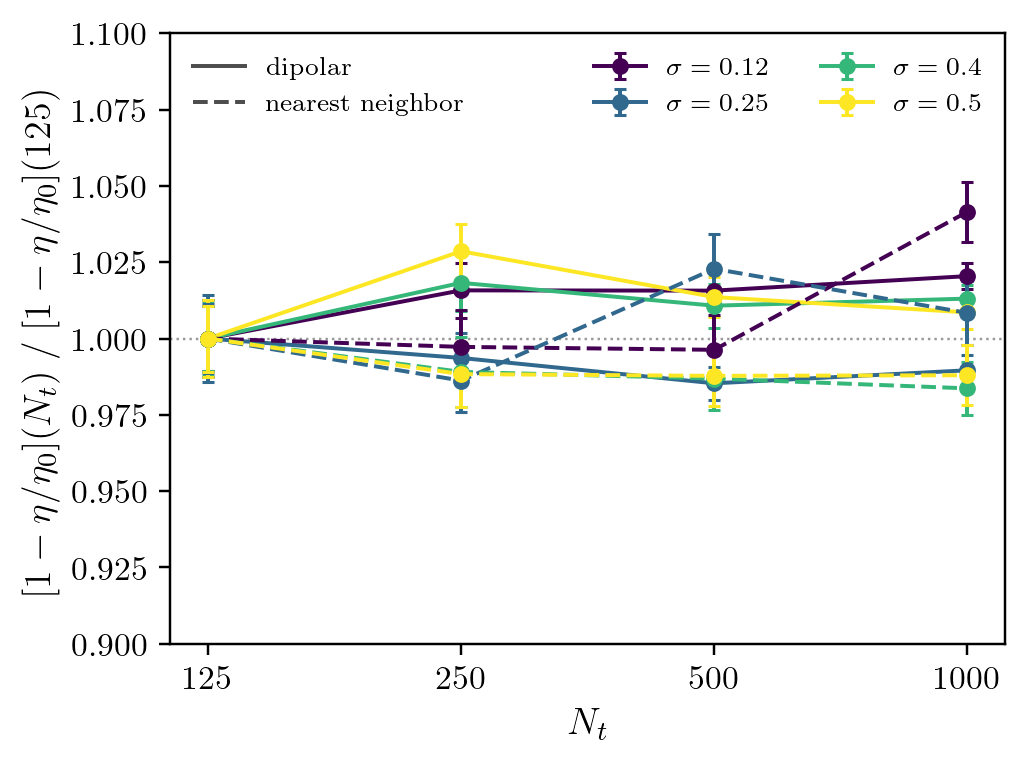}
\caption{Suppression $1-\eta(\sigma)/\eta(0)$ against $N_t$, normalized to its value at $N_t=125$, for four $\sigma$; dipolar solid, n.n. dashed, $V=0.3$. Antithetic sampling as described in the text.}
\label{fig:eta_N}
\end{figure}

At that cutoff the nearest-neighbor suppression is a small difference of two large numbers, and the scatter of single realizations is dominated by a first-order term. Since $\eta$ is translation invariant on the uniform ring, $\eta(\sigma)-\eta(0)=c_1\sum_iu_i+O(u^2)$, which averages to zero but is of order $\sigma\sqrt{N_t}$ in each realization, against a signal of order $\sigma^2$; in Eq.~\eqref{eq:tail_general} it is carried by $n_in_j-n_0^2\simeq n_0^2(u_i+u_j)$. We remove it by antithetic sampling, pairing each profile $n_i$ with its mirror $2n_0-n_i$, equally likely under the distribution of Eq.~\eqref{eq:halfwidth} and positive, which cancels all odd orders in $u_i$ exactly; for nearest neighbors at $\sigma=0.25$ this reduces the statistical error sixfold at fixed cost. The size scan of Fig.~\ref{fig:eta_N} uses these pairs, with the realization count raised at small $N_t$, to $400$ pairs at $N_t=125$, where diagonalization is cheap. The suppression is then independent of $N_t$ to within a few percent for both kernels between $N_t=125$ and $N_t=1000$. For the dipolar kernel at $N_t=1000$ it is also independent of the coupling to within $1\%$ across $V=0.3$, $0.1$ and $0.05$, at $\sigma=0.25$ and $0.5$, while $\xiloc$ changes by more than an order of magnitude.

\clearpage

\bibliographystyle{apsrev4-2}
\bibliography{dBECpaper}

\end{document}